\documentclass[fleqn,10pt,twocolumn]{wlscirep}
\usepackage[utf8]{inputenc}
\usepackage[T1]{fontenc}
\usepackage[normalem]{ulem}
\usepackage{authblk}

\usepackage{amsmath}
\usepackage{braket}
\usepackage{cleveref}
\usepackage{caption}
\usepackage{mathtools} 
\usepackage{multirow}
\usepackage{pgfplots}
\usepackage{subcaption}
\usepackage{tikz}
\usetikzlibrary{backgrounds,calc,shapes,arrows,positioning,fit}

\usepackage[acronym]{glossaries}
\glsdisablehyper  
\newacronym{bo}{BO}{Bayesian optimization}
\newacronym{cdft}{cDFT}{constrained DFT}
\newacronym{crpa}{cRPA}{constrained random-phase approximation}
\newacronym{dft}{DFT}{density-functional theory}
\newacronym{dfpt}{DFPT}{density-functional perturbation theory}
\newacronym{irrep}{irrep}{irreducible representation}
\newacronym{ks}{KS}{Kohn-Sham}
\newacronym{lr}{LR}{linear-response}
\newacronym{ml}{ML}{machine learning}
\newacronym{ocv}{OCV}{open-circuit voltage}
\newacronym{re}{RE}{rare-earth}
\newacronym{tm}{TM}{transition-metal}
\newacronym{rmse}{RMSE}{root mean square error}
\newacronym{sie}{SIE}{self-interaction error}
\newacronym{xc}{xc}{exchange-correlation}
\newacronym{lsda}{LSDA}{local spin-density approximation}
\newacronym{gga}{GGA}{generalized-gradient approximation}
\newacronym{sgga}{$\sigma$-GGA}{spin-polarized generalized-gradient approximation}
\newacronym{exx}{EXX}{exact exchange}
\newacronym{os}{OS}{oxidation state}
\newacronym{enn}{ENN}{equivariant neural-network}
\newacronym{scf}{SC}{self-consistent}
\newacronym{bz}{BZ}{Brillouin zone}
\newacronym{si}{SI}{supplemental information}
\newacronym{mare}{MARE}{mean absolute relative error}
\newacronym{sc}{SC}{self-consistent}
\newacronym{qe}{QE}{Quantum ESPRESSO}

\makeglossaries

\definecolor{mpink}{HTML}{BF016B}
\definecolor{mblack}{HTML}{222222}
\definecolor{mgray}{HTML}{CBCCC6}

\def\QE{\textsc{Quantum ESPRESSO}\,}

\newcommand{\editor}[2]{ 
  \expandafter\newcommand\csname #1note\endcsname[1]{%
    \textcolor{#2}{(\textbf{#1:} ##1)}}%
  \expandafter\newcommand\csname #1\endcsname[1]{%
    \textcolor{#2}{##1}}%
  \expandafter\newcommand\csname #1cancel\endcsname[1]{%
    \textcolor{#2}{\sout{##1}}}%
  \expandafter\newcommand\csname #1change\endcsname[2]{%
    \textcolor{#2}{\sout{##1} ##2}}%
  \newenvironment{#1text}{\color{#2}}{\color{black}}
}

\editor{MU}{blue}
\editor{IT}{green}
\editor{AZ}{teal}
\editor{LB}{magenta}

\title{Machine learning Hubbard parameters with equivariant neural networks
}

\author[1,2,$\dagger$]{Martin Uhrin}
\author[1]{Austin Zadoks}
\author[1]{Luca Binci\thanks{Present address: Department of Materials Science \& Engineering, University of California Berkeley, Berkeley, CA, 94720, USA; Materials Sciences Division, Lawrence Berkeley National Laboratory, Berkeley, CA, 94720, USA}}
\author[1,5]{Nicola Marzari}
\author[1,5,+]{Iurii Timrov}
\affil[1]{Theory and Simulation of Materials (THEOS), and National Centre for Computational Design and Discovery of Novel Materials (MARVEL), \'Ecole Polytechnique F\'ed\'erale de Lausanne (EPFL), CH-1015 Lausanne, Switzerland}
\affil[2]{Université Grenoble Alpes, 1130 Rue de la Piscine, BP 75, 38402 St Martin D'Heres, France}
\affil[5]{Laboratory for Materials Simulations (LMS), Paul Scherrer Institut (PSI), CH-5232 Villigen PSI, Switzerland}

\affil[$\dagger$]{Email: martin.uhrin@grenoble-inp.fr}
\affil[+]{Email: iurii.timrov@psi.ch}

\begin{abstract}
Density-functional theory with extended Hubbard functionals (DFT+$U$+$V$) provides a robust framework to accurately describe complex materials containing transition-metal or rare-earth elements.
It does so by mitigating self-interaction errors inherent to semi-local functionals which are particularly pronounced in systems with partially-filled d and f electronic states.
However, achieving accuracy in this approach hinges upon the accurate determination of the on-site $U$ and inter-site $V$ Hubbard parameters.
In practice, these are obtained either by semi-empirical tuning, requiring prior knowledge, or, more correctly, by using predictive but expensive first-principles calculations.
Here, we present a machine learning model based on equivariant neural networks which uses atomic occupation matrices as descriptors, directly capturing the electronic structure, local chemical environment, and oxidation states of the system at hand.
We target here the prediction of Hubbard parameters computed self-consistently with iterative linear-response calculations, as implemented in density-functional perturbation theory (DFPT), and structural relaxations.
Remarkably, when trained on data from 12 materials spanning various crystal structures and compositions, our model achieves mean absolute relative errors of 3\% and 5\% for Hubbard $U$ and $V$ parameters, respectively.
By circumventing computationally expensive DFT or DFPT self-consistent protocols, our model significantly expedites the prediction of Hubbard parameters with negligible computational overhead, while approaching the accuracy of DFPT.
Moreover, owing to its robust transferability, the model facilitates accelerated materials discovery and design via high-throughput calculations, with relevance for various technological applications.
\end{abstract}

\begin{document}

\flushbottom
\maketitle

\thispagestyle{empty}

\section*{INTRODUCTION}

A fundamental tool in investigating compounds involving \gls{tm} and \gls{re} compounds is \gls{dft},~\cite{Hohenberg:1964, Kohn:1965} a cornerstone for first-principles simulations in physics, chemistry, and materials science.
In practical applications, \gls{dft} necessitates approximations to the \gls{xc} functional, with the \gls{lsda} and \gls{sgga} being the most prevalent choices.
However, these approximations yield unsatisfactory outcomes for various properties of \gls{tm} and \gls{re} compounds, primarily due to significant \glspl{sie}~\cite{Perdew:1981, Kulik:2006, MoriSanchez:2006} that are particularly pronounced for localized d and f electrons.
To address these challenges, more accurate approaches surpassing the limitations of ``standard DFT'' have been devised.
Noteworthy among these are Hubbard-corrected DFT (so-called DFT+$U$~\cite{anisimov:1991, Liechtenstein:1995, Dudarev:1998} and its extension DFT+$U$+$V$,~\cite{Campo:2010, TancogneDejean:2020, Lee:2020} whose role in addressing \glspl{sie}, rather than correlation errors, was first pointed out in Ref.~\citenum{Kulik:2006}), meta-GGA functionals,~\cite{Sun:2015, Bartok:2019, Furness:2020} and hybrid functionals.~\cite{Adamo:1999, Heyd:2003, Heyd:2006}
While these methods offer valuable insights, each comes with inherent limitations and challenges, as discussed in, e.g., Ref.~\citenum{Timrov:2022b}.
Hubbard-corrected \gls{dft} in particular stands out for its greater accuracy with only a marginal increase in computational cost over standard \gls{dft} functionals.~\cite{Himmetoglu:2014}
In the DFT+$U$+$V$ scheme, a corrective Hubbard energy $E_{U+V}$ is introduced alongside the approximate DFT energy $E_{\mathrm{DFT}}$:
\begin{equation}
    E_{\mathrm{DFT}+U+V} = E_{\mathrm{DFT}} + E_{U+V} .
    \label{eq:Edft_plus_uv}
\end{equation}
In the simplified rotationally-invariant formulation,~\cite{Dudarev:1998} the extended Hubbard correction energy for a manifold with angular momentum $\ell$ takes the form:~\cite{Cococcioni:2005}
\begin{multline}
    E_{U+V} = \frac{1}{2} \sum_{I \sigma} U^I \mathrm{Tr} \left[ \mathbf{n}_{\ell}^{II\sigma} \left( \mathbf{1} - \mathbf{n}_{\ell}^{II\sigma} \right) \right] \\
    - \frac{1}{2} \sum_{I} \sum_{J (J \ne I)}^{*} \sum_{\sigma} V^{IJ} \mathrm{Tr} \left[ \mathbf{n}^{IJ\sigma}_{\ell} \mathbf{n}^{JI\sigma}_{\ell} \right] \,,
    %
    \label{eq:Edftu}
\end{multline}
where $I$ and $J$ are atomic site indices, and $\sigma$ is the spin index.
$U^I$ and $V^{I J}$ are effective on-site and inter-site Hubbard parameters, respectively.
The asterisk in the sum signifies that, for each atom $I$, the index $J$ encompasses all its neighbors up to a given distance.
The generalized occupation matrices $\mathbf{n}^{I J \sigma}_{\ell}$ are derived from the projection of the \gls{ks} states onto localized atom-centered orbitals $\phi^{I}_{m}(\mathbf{r})$ (Hubbard projector functions) of neighboring atoms:
\begin{equation}
    \mathbf{n}_{\ell}^{IJ\sigma} \equiv n^{I J \sigma}_{m m'} = \sum_{v,\mathbf{k}} f^\sigma_{v,\mathbf{k}}
    \braket{\psi^\sigma_{v,\mathbf{k}} | \phi^{J}_{m'}} \braket{\phi^{I}_{m} | \psi^\sigma_{v,\mathbf{k}}} \,,
    \label{eq:occ_matrix_0}
\end{equation}
where $v$ represents the band index of the \gls{ks} wavefunctions $\psi^\sigma_{v,\mathbf{k}}(\mathbf{r})$, $m$ and $m'$ denote magnetic quantum numbers associated with a specific angular momentum $\ell$, $\mathbf{k}$ denotes points in the first \gls{bz}, and $f^\sigma_{v,\mathbf{k}}$ are the occupations of the \gls{ks} states.
The two terms in \cref{eq:Edftu} — proportional to on-site $U^{I}$ and inter-site $V^{IJ}$ — counteract one another.
The on-site term promotes localization on atomic sites, suppressing hybridization with neighbors, while the inter-site term favors hybridized states with components on neighboring atoms.
Consequently, the values of $U^I$ and $V^{IJ}$ are critical to optimizing the extent of the localization and hybridization within Hubbard-corrected \gls{dft}.
However, these parameters are not known \textit{a priori} and must be determined in some way.
In passing we note that for the sake of simplicity, hereafter we drop the superscripts $I$ and $J$ in the notations of Hubbard parameters and occupation matrices, unless required for clarity.

Hubbard $U$ can be fit semi-empirically to reproduce a target property from experimental data~\cite{Wang:2006, Kubaschewski:1993, LeBacq:2004, Aykol:2014, Isaacs:2017} or from other advanced first-principles methods (e.g. $GW$~\cite{Hedin:1965} or hybrids~\cite{Adamo:1999, Heyd:2003, Heyd:2006}); however, this approach has many limitations.
Fitted parameters are not guaranteed to, and often do not, accurately predict properties other than those used for fitting, and calibrating a single Hubbard $U$ to multiple properties is non-trivial, requiring advanced algorithms like \gls{bo}.~\cite{Yu:2020, Tavadze:2021}
This method also precludes materials discovery, where properties are by definition unknown, and targeting results of advanced first-principles methods, like DFT with hybrid functionals, inherits the limitations of those methods.~\cite{Skone:2014,Skone:2016}
Moreover, the inter-site $V$ parameters, which are necessary to properly describe materials with pronounced covalent interactions, are difficult to fit semi-empirically due to the high-dimensional regression procedure they would require.
An attractive alternative is computing Hubbard parameters using first-principles methods such as \gls{cdft},~\cite{Dederichs:1984, Mcmahan:1988, Gunnarsson:1989, Hybertsen:1989, Gunnarsson:1990, Pickett:1998, Solovyev:2005, Nakamura:2006, Shishkin:2016} Hartree-Fock-based approaches,~\cite{Mosey:2007, Mosey:2008, Andriotis:2010, Agapito:2015, TancogneDejean:2020, Lee:2020} and the \gls{crpa}.~\cite{Springer:1998, Kotani:2000, Aryasetiawan:2004, Aryasetiawan:2006}
These methods do not rely on data from experiments or advanced simulation methods and, moreover, are able to provide values not only for $U$ but also for $V$.
However, first-principles approaches are considerably more computationally expensive than \gls{dft}+$U$(+$V$) ground-state calculations themselves, making their application feasible but demanding for large systems or high-throughput studies.

\begin{figure*}[ht]
    \includegraphics[width=15cm]{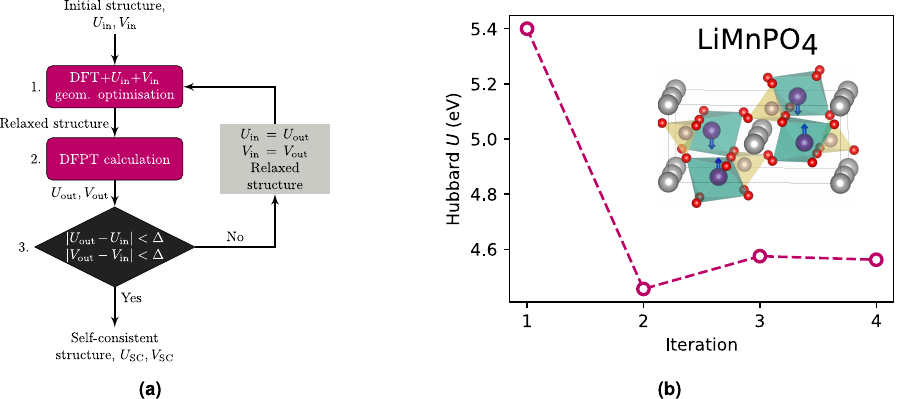}
    \caption{\textbf{Calculating Hubbard corrections self-consistently using density-functional perturbation theory.}
    a) Protocol for the self-consistent calculation of Hubbard parameters using \gls{dfpt}.~\cite{Timrov:2021} $U_\mathrm{in}$ and $V_\mathrm{in}$ represent the input Hubbard parameters, while $U_\mathrm{out}$ and $V_\mathrm{out}$ denote the output parameters, with $\Delta$ representing the convergence threshold. $U_\mathrm{SC}$ and $V_\mathrm{SC}$ are the final \gls{scf} Hubbard parameters.
    b) Convergence of the Hubbard $U$ parameter for Mn-3d states in LiMnPO$_4$ using the self-consistent protocol.\cite{Timrov:2022b}
    The inset displays the crystal structure of the material, where arrows indicate the spin direction, and Li atoms are depicted in grey, O in red, Mn in violet, and P in yellow.}
    \label{fig:self-consistent}
\end{figure*}

The \gls{lr} formulation of \gls{cdft} (\gls{lr}-\gls{cdft})~\cite{Cococcioni:2005} has witnessed popularity due to its simplicity and accuracy; however, it demands computationally expensive supercell calculations. A recent reformulation of \gls{lr}-\gls{cdft} in terms of \gls{dfpt}~\cite{Timrov:2018, Timrov:2021} significantly reduces the computational burden for determining Hubbard parameters by replacing cumbersome supercell calculations with faster unit-cell calculations and by leveraging symmetries that further diminishes the computational cost by reducing the number of perturbations in reciprocal space (see Sec.~S1 in the \gls{si}).
Its physical rationale relies on heuristically imposing piecewise linearity of the total energy of the system as a function of the population of the Hubbard manifold.~\cite{Cococcioni:2005}
Despite the numerous successful applications of \gls{dfpt} in computing Hubbard parameters,~\cite{Timrov:2020c, Mahajan:2022, Timrov:2022b, Timrov:2023, Binci:2023, Haddadi:2023, Bonfa:2023, Gelin:2023, Macke:2023} this approach introduces a significant overhead compared to \gls{dft}+$U$+$V$ ground-state calculations.
Moreover, it has been shown that jointly optimizing Hubbard parameters and the crystal structure, rather than relying on the equilibrium geometry obtained using (semi-)local functionals, can significantly improve the accuracy of the final properties of interest.~\cite{Timrov:2020b}
This is due in part to taking into account the geometry dependence of $U\{\mathbf{R}\}$ and thus the contribution $dU/d\mathbf{R}$ to the Hellmann-Feynman forces.~\cite{Kulik:2011b}
To do so, a self-consistent procedure combining \gls{dfpt} and structural optimizations (\cref{fig:self-consistent}) can be used (see Sec.~S2 in the \gls{si}).~\cite{Hsu:2009, Timrov:2021}
\Cref{fig:self-consistent} shows a typical self-consistent determination of Hubbard parameters for LiMnPO$_4$ following this procedure, where each iteration costs approximately one order of magnitude more computational time than a Hubbard-corrected \gls{dft} electronic ground-state calculation.
Consequently, the adoption of acceleration techniques is highly sought after to expedite the determination of Hubbard parameters while providing a level of accuracy very close to that of \gls{dfpt} and, prospectively, a simple estimate of $dU/d\mathbf{R}$ and $dV/d\mathbf{R}$.

Using \gls{ml} to accelerate parts of a DFT workflow is becoming increasingly common,~\cite{DeMendonca2023,Schubert2024} including applications to predicting Hubbard parameters.
A recent example is the work of Cai et al.~\cite{Cai:2024}, in which they use crystal structure parameters, such as bond lengths and angles (among others), as descriptors for a random forest regression model trained on a database of Mn oxides.
Their training data consist of $U$ values fitted using Bayesian optimization to replicate band gaps and band structures obtained from HSE hybrid functional calculations — a concept initially proposed in the earlier work of Ref.~\citenum{Yu:2020} within the context of DFT+$U$ and subsequently extended to the DFT+$U$+$V$ framework.~\cite{Yu:2023}
The resulting \gls{ml} model predicts $U$ values such that subsequent DFT+$U$ calculations using these values yield band gaps that are $2-3$ times smaller than the reference HSE values, while the band structures appear qualitatively similar.
Despite the significance of these efforts in applying \gls{ml} to the prediction of Hubbard parameters, numerous critical issues persist, such as the accuracy of the reference training data (HSE often provides unreliable band gaps for solids~\cite{Skone:2014,Skone:2016}), the choice of the target properties for the training data (fitting $U$ to reproduce band gaps is questionable), the importance of self-consistency between Hubbard parameters and the crystal structure, as well as the lack of inter-site $V$, to name a few.
Addressing these points is highly relevant and necessitates further investigation, which serves as the motivation for this work.

Here, we introduce a novel \gls{ml} approach based on \glspl{enn} which aims to replace computationally demanding first-principles \gls{dfpt} calculations of Hubbard $U$ and $V$ parameters while providing a negligible loss of accuracy for the vast majority of practical applications.
Crucially, the model employs: 1)~atomic occupation matrices within the DFT+$U$+$V$ framework as descriptors of the geometry, electronic structure (e.g. \glspl{os}) and local chemical environments in materials, as well as 2)~\gls{dfpt}-based Hubbard parameters, and 3)~interatomic distances.
This model is general and can be applied to materials with ionic, covalent, and mixed ionic-covalent interactions.
It is trained on all intermediate Hubbard parameters and occupation matrices obtained during the self-consistent cycle and directly provides the final self-consistent values of $U$ and $V$, thus also bypassing intermediate structural optimizations.
The utilization of \glspl{enn} facilitates the exploitation of the inherent O(3) group structure of the occupation matrices, ensuring excellent model performance even with scarce training data.
Equivariant models have demonstrated state-of-the-art accuracy and transferability in \gls{ml} interaction potentials,~\cite{Batzner2021, Musaelian2022} while our work is the first to incorporate electronic-structure degrees of freedom as explicit features in solids.

\section*{RESULTS}

\paragraph{Occupation matrices as the model inputs}
The goal of our \gls{ml} approach is to replace the self-consistent procedure represented in \cref{fig:self-consistent} and provide the final Hubbard parameters using as input results from an initial DFT(+$U$+$V$) calculation.
Conventional \gls{ml} methods for atomistic systems primarily use the ionic structures as inputs (see e.g.~Refs.~\citenum{Bartok2013a,Musil2021,Uhrin2021b,Batatia2023}), side-stepping the need to explicitly calculate or even consider the electronic structure.
However, as a consequence such models may not be particularly sensitive to changes in electronic structure that only lead to subtle changes in local atomic geometry, as can happen during a change of \gls{os}.
To overcome this limitation, our models take as input the on-site occupation matrices $\mathbf{n}^\sigma_{\ell} \equiv \mathbf{n}_{\ell}^{II \sigma}$ [see  \cref{eq:occ_matrix_0}, $I = J$], which describe the local electronic structure around an atom, including the \gls{os} which is reflected in the occupation matrix eigenvalues.~\cite{Sit:2011}
This is a particularly compelling choice given the significant variations in Hubbard parameters seen for different \glspl{os} of TM elements.~\cite{Timrov:2022b, Timrov:2023} We emphasize that relying solely on the trace of the occupation matrix as a descriptor may not adequately distinguish between different OSs and chemical environments.

As with any learning task, the \gls{ml} model should respect the way in which the target physical properties transform under global rotations, translations, reflections and permutations of labels.
In the present case, the output $U$ and $V$ values are left unchanged by these transformations, however the entries of the occupation matrix \emph{do} change under rotation, making $n^\sigma_{m m^\prime}$ unsuitable for use as inputs to non-symmetry aware models.
For this reason, we make use of an \emph{equivariant} learning model,~\cite{Kondor2018c,Thomas2018,Grisafi2019,Geiger2021} which has the property that the learned function $f: X \to Y$ obeys the following relation:
\begin{equation}
    D_Y[g]f(x) = f(D_X[g]x), \quad \forall g \in G, \,\, \forall x \in X,
\end{equation}
where the equivariance is with respect to a group $G$ (in our case SE(3)), and $D_Y[g]$ and $D_X[g]$ are representations that act on the vector spaces $Y$ and $X$, respectively.
For example, applying a rotation to the inputs and then applying $f$ must produce the same outputs as applying the rotation to the outputs of $f$ given the original (unrotated) input.
While our model outputs scalars (which are symmetry invariant), by using an equivariant model we ensure that the inputs and all intermediate (hidden) features transform together under group actions.
This property has been shown to give state-of-the-art accuracy and transferability, particularly within the domain of \gls{ml} interaction potentials.~\cite{Batzner2021,Merchant2023,Batatia2023}

\paragraph{Equivariant descriptors}
By virtue of being atom-centred, the on-site occupation matrix $\mathbf{n}^\sigma_{\ell}$ is naturally invariant to global translations.
However, to build a rotationally invariant model, it is convenient to re-express the occupation matrix in terms of the \glspl{irrep} of the O(3) group.
If we let $\Gamma^{(\ell,P)}$ be the \gls{irrep} with degree $\ell$ and parity $P$, we can express the occupation matrices as the tensor products $\Gamma^{(\ell,P)} \otimes \Gamma^{(\ell,P)}$, where e.g. $\ell = 1$ for p orbitals, and $\ell = 2$ for d orbitals.
These can then be decomposed into a direct sum of \glspl{irrep}, e.g. for $\ell = 1$ and $P = -1$ the occupation matrix is a $3 \times 3$, rank-2, tensor $\Gamma^{(1,-1)} \otimes \Gamma^{(1,-1)}$, that is symmetric, as $\mathbf{n}^\sigma_{\ell} = \left( \mathbf{n}^\sigma_{\ell} \right)^\intercal$.
In practice, this decomposition is achieved by applying a change of basis for $\mathbf{n}^\sigma_{\ell}$ that transforms it into the \gls{irrep} basis (see Supplementary Section 3 for details).
At this stage, the remaining symmetry to be addressed is that of label permutation.
In the case of the atom labels for the inter-site term, the model should be invariant to permutation of $I$ and $J$, however in our training data the Hubbard $V$ correction is always applied to a d-block/p-block atom pairs which have occupation matrices of different dimension ($5 \times 5$ for d and $3 \times 3$ for p), thereby making them distinguishable.

\begin{table}[t]
    \renewcommand{\arraystretch}{1.3}
    \centering
    \begin{tabular}{cl}
        \hline\hline
        Attribute                      & Irreducible representation                                   \\
        \hline
        $U^I$, $V^{IJ}$, $r_{IJ}$      & $\Gamma^{(0,1)}$                                             \\
        Atomic species                 & $\bigoplus^{N_s} \Gamma^{(0,1)}$                             \\
        $\mathbf{n}^\sigma_\mathrm{p}$ & $\Gamma^{(0,1)} \oplus \Gamma^{(2,1)}$                       \\
        $\mathbf{n}^\sigma_\mathrm{d}$ & $\Gamma^{(0,1)} \oplus \Gamma^{(2,1)} \oplus \Gamma^{(4,1)}$ \\
        \hline\hline
    \end{tabular}
    \caption{Irreducible representations of various attributes (inputs and outputs) of the ML model. $U^I$ and $V^{IJ}$ are the on-site and inter-site Hubbard parameters, respectively, $r_{IJ}$ is the interatomic distance between species $I$ and $J$, $N_s$ is the number of atomic species (one-hot vectors), $\mathbf{n}^\sigma_\mathrm{p}$ and $\mathbf{n}^\sigma_\mathrm{d}$ are the occupation matrices for the p and d orbitals, respectively, in the special representation for the ML model [see~\cref{eq:occ_1,eq:occ_2}]. The irreducible representations of the occupation matrices can be readily obtained using the \texttt{e3nn}~\cite{Geiger2021,Geiger2022} library.}
    \label{tab:representations}
\end{table}

There remains a permutational invariance to be imposed on the spin labels ($\sigma = \uparrow$ or $\downarrow$).
We achieve this using permutationally invariant polynomials:
\begin{align}
    \mathbf{x}^1_{\ell} & = \mathbf{n}_{\ell}^{\uparrow} + \mathbf{n}_{\ell}^{\downarrow} \,, \label{eq:occ_1}       \\
    \mathbf{x}^2_{\ell} & = \mathbf{n}_{\ell}^{\uparrow} \otimes \mathbf{n}_{\ell}^{\downarrow} \,. \label{eq:occ_2}
\end{align}
In the following, for the sake of convenience, we label these tensors in their \gls{irrep} form as $\mathbf{x}^i_{\mathrm{p}}$ for p orbitals and $\mathbf{x}^i_{\mathrm{d}}$ for d orbitals, where $i$ can be 1 or 2.
We note in passing that in the non-spin-polarized case, one can set $\mathbf{n}_\ell^\uparrow = \mathbf{n}_\ell^\downarrow = \mathbf{n}_\ell/2$, where $\mathbf{n}_\ell$ is divided by two to account for spin degeneracy, and then proceed as above.
In addition, atomic species (i.e. atomic types) are encoded as one-hot vectors in our \gls{ml} model.
A summary of the \glspl{irrep} of all the model inputs is presented in \cref{tab:representations}.

In the case of the \gls{ml} model for Hubbard $V$, the interatomic distances $r_{IJ}$ are also included as an additional input, giving some basic information on the ionic structure.
It is important to note that we exclusively focus on the on-site occupation matrices $\mathbf{n}_{\ell}^{II\sigma}$ for practical reasons.
While inter-site matrices (see e.g.~\cite{Nigam2021a, Zhang2022}) could be naturally included in our models, Quantum ESPRESSO does not currently output these.
Additionally, this approach ensures consistency with the linear-response approach used to calculate the Hubbard $V$ parameters from first principles, which relies on the response of on-site occupation matrices only~\cite{Campo:2010, Timrov:2021}.
However, using inter-site occupation matrices $\mathbf{n}_{\ell}^{IJ\sigma}$ as descriptors instead of $r_{IJ}$ presents an interesting alternative that could be explored in future studies.
Finally, while additional descriptors, such as the derivatives of the occupation matrix with respect to the Hubbard parameters, could potentially enhance the performance of the ML model, this would significantly increase the cost of generating training data.

\begin{figure*}[ht]
    \centering
    \includegraphics[width=16cm]{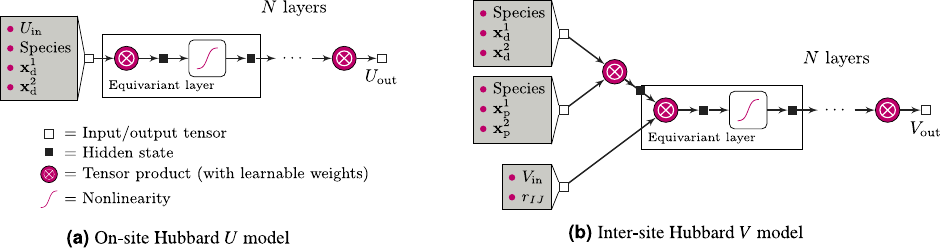}
    \caption{\textbf{Schematic illustration of the equivariant neural-network ML model for predicting $U$ and $V$ Hubbard parameters.} $U_\mathrm{in}$ and $V_\mathrm{in}$ are the input Hubbard parameters, while $U_\mathrm{out}$ and $V_\mathrm{out}$ are the outputs. The atomic species enter as one-hot tensors, and $r_{IJ}$ is the interatomic distance between sites $I$ and $J$. $\mathbf{x}^1_\mathrm{d}$ and $\mathbf{x}^2_\mathrm{d}$ are the occupation matrices for the d orbitals, while $\mathbf{x}^1_\mathrm{p}$ and $\mathbf{x}^2_\mathrm{p}$ for the p orbitals, all in the special representation for the ML model [see~\cref{eq:occ_1,eq:occ_2}]. Tensors are represented as squares: open for inputs and outputs, and filled for intermediate features.}
    \label{fig:ml_models}
\end{figure*}

\paragraph{Equivariant neural-network model}
To construct our \gls{ml} model, we use the \texttt{e3nn} library~\cite{Geiger2021, Geiger2022} and PyTorch;~\cite{Paszke2019} our codebase is open-source and freely available (see \textit{code availability} below).
As shown in \cref{fig:ml_models}, we define two separate models, one for predicting the on-site $U$ values and the other for the inter-site $V$ ones.
By using separate models, we provide flexibility for the model to be used in calculations where only Hubbard $U$ is applied (i.e. DFT+$U$).
In each case, the model starts with one or more nodes that carry attributes, expressed as a direct sum of \glspl{irrep}, which represent the inputs to the learned function.
The inputs pass through a series of repeated layers made up of a tensor product (analogous to all-to-all connected layers in a traditional neural network) and a gated non-linearity (an equivariant version of a traditional activation function~\cite{Geiger2022}) followed by a final tensor product before readout.
For each experiment, we rescale the model outputs to have zero mean and a standard deviation of one based on the Hubbard parameters found in the training set which simultaneously accelerates training and improves the final loss.
All of the tensor products have learnable weights, meaning that every pair of input \glspl{irrep} that contribute to one output \gls{irrep} has a learnable scalar parameter that is optimised during training.
We use the AdamW~\cite{Loshchilov2017} optimiser to minimise the loss, which is the mean squared error between predicted and training Hubbard parameters.

\paragraph{Training and validation datasets}
To train and validate our \gls{enn} model, we curated a dataset comprising materials with diverse crystal structures.
This dataset is constructed based on an investigation of various Li-ion cathode materials, covering olivine-type,~\cite{Timrov:2022b} spinel-type,~\cite{Timrov:2023} and layered-type~\cite{Chakraborty:2018} structures at different Li concentrations, employing the self-consistent DFT+$U$ and DFT+$U$+$V$ protocols outlined above.
Additionally, we include materials such as tunnel- and rutile-type MnO$_2$~\cite{Mahajan:2021, Mahajan:2022} and perovskite-type rare-earth nickelates~\cite{Binci:2023} in our dataset.
The crystal and electronic structure details, magnetic ordering, and various other properties of these materials are extensively documented in their respective publications.
These are the compounds containing \gls{tm} elements Ni, Mn, and Fe, and a list of all these materials is provided in \cref{tab:materials}.
While our current dataset is relatively small in terms of material count, ongoing efforts are directed towards a high-throughput exploration involving hundreds of materials, which will serve as an expanded training and validation set for a future, general-purpose, model.
This initial dataset demonstrates the efficacy of our \gls{ml} model, showcasing its ability to predict Hubbard parameters even with limited data (see Table~\ref{tab:training_data}).
Despite the size of the dataset, the model exhibits accurate predictions, setting the stage for further refinement and expansion with the upcoming comprehensive investigation.

\begin{table}[t]
    \centering
    \begin{tabular}{l l r}
        \hline\hline
        Crystal structure type      & Chemical composition             & \# \\
        \hline
        \multirow{3}{*}{Olivine}    & Li$_x$FePO$_4$                   & 5  \\
                                    & Li$_x$MnPO$_4$                   & 5  \\
                                    & Li$_x$Fe$_{0.5}$Mn$_{0.5}$PO$_4$ & 5  \\
        \hline
        \multirow{2}{*}{Spinel}     & Li$_x$Mn$_2$O$_4$                & 2  \\
                                    & Li$_x$Mn$_{1.5}$Ni$_{0.5}$O$_4$  & 2  \\
        \hline
        \multirow{2}{*}{Layered}    & Li$_x$NiO$_2$                    & 2  \\
                                    & Li$_x$MnO$_2$                    & 2  \\
        \hline
        Tunnel                      & Fe-doped $\alpha$-MnO$_2$        & 1  \\
        Tunnel                      & $\alpha$-MnO$_2$                 & 1  \\

        Rutile                      & $\beta$-MnO$_2$                  & 1  \\
        \hline
        \multirow{2}{*}{Perovskite} & YNiO$_3$                         & 1  \\
                                    & PrNiO$_3$                        & 1  \\
        \hline
        Total                       &  12                              & 28 \\
        \hline\hline
    \end{tabular}
    \caption{A list of the materials that are used to train and validate the ML model. The last column shows the total number of materials. For the olivine-type materials $x = 0, 0.25, 0.50, 0.75, 0$, while for the spinel- and layered-type materials $x = 0, 1$.}
    \label{tab:materials}
\end{table}

For each material in our dataset, \gls{scf} Hubbard $U$ and $V$ parameters are computed using the self-consistent protocol illustrated in \cref{fig:self-consistent}.
We include all $V$ parameters for Hubbard-active atom pairs whose \gls{dfpt} values are greater than $0.3$~eV, in in constrast to Ref.~\citenum{Yu:2023} where only $V$ for the nearest neighbor couples were included in the training of their \gls{ml} model.
Throughout the iterative process, all intermediate converged occupation matrices and Hubbard parameters are systematically saved.
The dataset for training $V$ comprises approximately seven times more data points than that for $U$ as $V$ is an atom-pair quantity, in contrast to the local and consequently sparser nature of $U$.
For each material and chemical element, this data is randomly split, holding back 20\% from each material for validation.

\begin{table}[t]
    \centering
    \begin{tabular}{l c c}
        \hline\hline
        Chemical element & \multicolumn{2}{c}{Number of data points}                     \\
                         & $U$                                       & $V$               \\
        \hline
        Ni               & 396 (124)                                 & 67,802 (63,332)   \\
        Mn               & 856 (284)                                 & 162,272 (153,232) \\
        Fe               & 138 (120)                                 & 22,511 (21,483)   \\
        \hline
        Total            & 1,390 (528)                               & 252,585 (238,047) \\
        \hline\hline
    \end{tabular}
    \caption{Chemical elements and the total number of data points for both on-site $U$ and inter-site $V$ Hubbard parameters determined using the \gls{scf} protocol illustrated in \cref{fig:self-consistent}. These data points are aggregated across all the materials listed in \cref{tab:materials} that contain the respective chemical element.
        Numbers in brackets indicate the unique data points after de-duplication (see text).}
    \label{tab:training_data}
\end{table}

In the first-principles calculations, certain input Hubbard parameters ($U_\mathrm{in}$ and $V_\mathrm{in}$) yield a specific ground state and occupation matrices, and the response of the system due to a perturbation through \gls{dfpt} provides corresponding output Hubbard parameters ($U_\mathrm{out}$ and $V_\mathrm{out}$).
We use these quantities to predict either the final self-consistent Hubbard parameter values, or the result of a single \gls{dfpt} calculation to test various aspects of the model (see details for each experiment below).
It is worth noting that many of the structures contain symmetry-equivalent sites, and correlations between results from successive self-consistent steps are prevalent.
To address this issue, we implement a de-duplication procedure aimed at mitigating potential leakage of identical data from the training to the validation set.
This involves calculating a symmetry-invariant distance between all pairs of \gls{enn} inputs for each atomic species in each material, which is then used to cluster duplicates based on a specified distance threshold (see Supplementary Section 4).
In random train/validate splits, we sample from these clusters rather than individual data points, ensuring that only one example from each cluster is included in the validation set.

\begin{figure*}[ht]
    \centering
    \includegraphics[width=15cm]{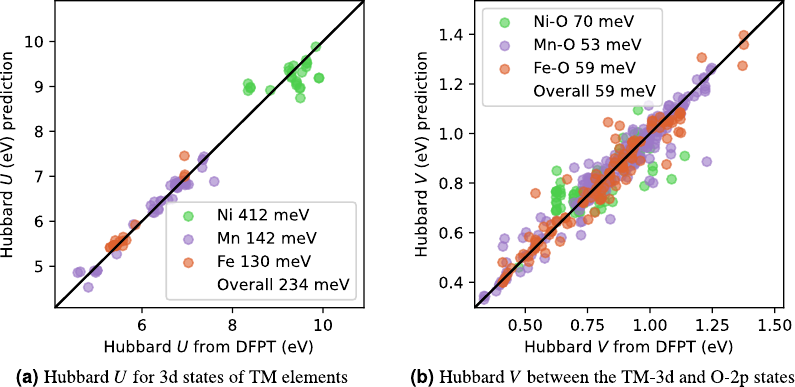}
    \caption{\textbf{Parity plots showing the prediction accuracy on an unseen validation dataset.} The energies in the legend are the RMSE categorized by element(s) and the overall RMSE across all elements. All attributes listed in~\cref{tab:representations} are used as inputs for the ML model.}
    \label{fig:predict_final}
\end{figure*}

\paragraph{Accuracy of the ML model in predicting the \gls{scf} Hubbard parameters}
First, we demonstrate the model's ability to directly learn the final \gls{scf} Hubbard parameters for the TM elements across all the materials in our dataset.
This is achieved using all attributes listed in~\cref{tab:representations} as inputs for our \gls{ml} model.
\Cref{fig:predict_final} shows the parity plots obtained for the on-site $U$ and inter-site $V$ Hubbard parameters.
The \gls{mare} over all species consistently remains below 3\% and 5\% for the $U$ and $V$ parameters, respectively (per-species distributions are reported in Sec.~S5 in the \gls{si}).
The result for Hubbard $U$ is particularly promising, considering the relatively small amount of available training data.
The higher \gls{mare} obtained for $V$ is likely due to the significantly increased number of degrees of freedom in its model, owing to the involvement of pairs of atoms.
It is worth noting that in order to attain this accuracy in predicting the $U$ and $V$ values, utilizing a relatively small batch size of $8-16$ was beneficial for preventing overfitting.

Second, we examine the extent to which our \gls{ml} model depends upon the input Hubbard parameters for its prediction accuracy. To investigate this, we repeat the same numerical experiment as above, eliminating the input Hubbard parameters ($U_\text{in}$ and $V_\text{in}$) from our model.
As detailed in Sec.~S6 in the \gls{si}, the parity plots closely resemble those in \cref{fig:predict_final}. The reduction in accuracy of our \gls{ml} model resulting from this simplification is relatively small, namely the overall \gls{rmse} is increases by 29 and 18~meV for the on-site $U$ and inter-site $V$, respectively. This finding suggests that the input Hubbard parameters may not be the most critical input attributes for our \gls{ml} model, and that the input occupation matrices contain sufficient information to make low-error predictions.

\begin{table}[t]
    \centering
    \begin{tabular}{clcc}
        \hline\hline
        $x$                    & Property      & Li$_x$MnPO$_4$ & Li$_x$FePO$_4$ \\
        \hline
        \multirow{3}{*}{$0-1$} & $\Phi$ (DFPT) & 4.205          & 3.544          \\
                               & $\Phi$ (ML)   & 4.194          & 3.544          \\
                               & $\Delta \Phi$ & $-0.26$\%      & 0.00\%         \\
        \hline
        \multirow{3}{*}{0}     & $m$ (DFPT)    & 3.9738         & 4.1828         \\
                               & $m$ (ML)      & 3.9720         & 4.1832         \\
                               & $\Delta m$    & $-0.05$\%      & $0.01$\%       \\
        \hline
        \multirow{3}{*}{1}     & $m$ (DFPT)    & 4.7482         & 3.7388         \\
                               & $m$ (ML)      & 4.7486         & 3.7391         \\
                               & $\Delta m$    & $0.01$\%       & $0.01$\%       \\
        \hline\hline
    \end{tabular}
    \caption{Comparison of the open-circuit voltages $\Phi$ (in V) and magnetic moments for TM elements (in $\mu_\mathrm{B}$) for Li$_x$MnPO$_4$ and Li$_x$FePO$_4$ computed within DFT+$U$+$V$, with $U$ and $V$ obtained from first principles using DFPT and predicted using the ML model. The voltages are computed using the total energy differences for the Li concentrations $x=0$ and $x=1$, while the magnetic moments are computed as the trace of the difference between the spin-up and spin-down occupation matrices.~\cite{Timrov:2022b} $\Delta \Phi$ and $\Delta m$ are the relative differences between the voltages and magnetic moments based on DFPT and ML, respectively.}
    \label{tab:ocvs}
\end{table}

In addition to evaluating the model accuracy, it is also important to evaluate its impact on downstream properties, such as voltages in Li-ion battery cathode materials.
\Cref{tab:ocvs} shows a comparison of calculated \gls{ocv}~\cite{Timrov:2022b} $\Phi$ and magnetic moments within the DFT+$U$+$V$ framework, using first-principles \gls{scf} Hubbard parameters from \gls{dfpt} and those predicted by our \gls{ml} model.
Such OCV have been calculated using standard definitions \cite{Aydinol:1997,Zhou:2004}, which, when employed in the present context, imply that the energies of different Li concentrations are evaluated using different Hubbard parameters. This approach was used and discussed in more detail in previous works~\cite{Cococcioni:2019, Timrov:2022b, Mahajan:2021}, while our goal here is to showcase the consistency between the results obtained with first-principles Hubbard parameters and the ones calculated through the ML model.
We find that the differences between the computed and predicted $U$ and $V$ values are less than $0.12$~eV for Li$_x$MnPO$_4$ and Li$_x$FePO$_4$ ($x=0$ and $x=1$).
Generally, such a small variation in the values of Hubbard parameters has a negligible impact on the vast majority of various physical and chemical properties of materials.
The differences observed in the \gls{ocv} and magnetic moments in \Cref{tab:ocvs} are indeed negligible, indicating the accuracy and reliability of the \gls{ml}-predicted Hubbard parameters.

\paragraph{ML model's performance using a reduced number of iterations in the \gls{scf} protocol}
In the preceding section, we assessed the \gls{ml} model's performance in predicting final self-consistent Hubbard parameters using training data that consists of a subset of \gls{dfpt} calculations from all iterations of the \gls{scf} protocol (see \cref{fig:self-consistent}).
However, this approach relies on conducting numerous computationally intensive \gls{dfpt} calculations to generate these training data (often $2-5$ but occasionally up to $10$ per material to reach self-consistency).

In this experiment, we instead investigate how well the model can predict the results of a {\it single} \gls{dfpt} calculation as a function of how many self-consistent iterations are performed in generating training data.
This task is more targeted towards high-throughput applications where data generation for model training could be more efficiently performed by sampling the first few \gls{dfpt} calculations in the self-consistent workflows of many materials rather than the highly-correlated \gls{dfpt} calculations of full workflows for a few materials.
Effectively, this approach aims to stop the self-consistent cycle early and use a model trained on the calculations already performed to refine the not-yet-self-consistent result to be closer to self-consistency, essentially for free.
We investigate this use case by training the model on only the first $N_\mathrm{iter} - 1$ \gls{dfpt} results, and asking it to predict the outcome of the next \gls{dfpt} calculation and study how the model performs for different values of $N_\mathrm{iter}$.

\begin{figure*}[ht]
    \centering
    \includegraphics[width=0.98\columnwidth]{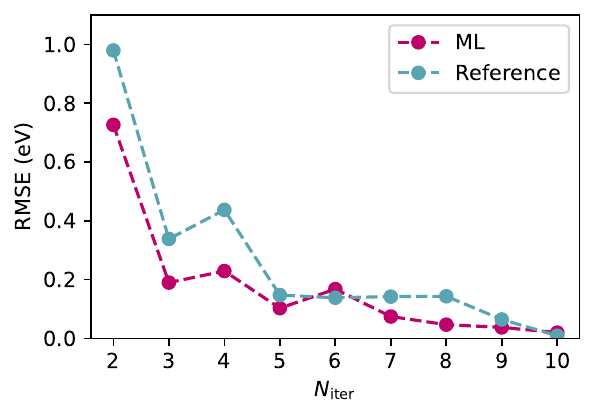}
    \caption{\textbf{Evaluation of model performance using a reduced number of self-consistent steps.}
    RMSE for all materials as a function of the number of iterations $N_\mathrm{iter}$ in the \gls{scf} protocol (see \cref{fig:self-consistent}).
        At each $N_\mathrm{iter}$ the plotted reference value is the \gls{rmse} between the \gls{dfpt} result obtained from the previous \gls{scf} iteration and the current one (i.e. a measure of how much the Hubbard $U$ changes by doing one more iteration).
        The \gls{ml} value shows the prediction made from having trained on results from all the preceding $N_\mathrm{iter} - 1$ iterations and predicting the $N_\mathrm{iter}$th value.
        This helps to answer the question of how much the \gls{ml} model can improve the Hubbard parameter, at essentially no cost, when limiting the number of \gls{dfpt} steps as might be done during the high-throughput screening.}
    \label{fig:iteration_comparison}
\end{figure*}

\Cref{fig:iteration_comparison} shows the average \gls{rmse} over all elements as the number of \gls{scf} iterations used for training $N_\mathrm{iter}$ increases.
The reference data correspond to \gls{rmse} evaluated using \gls{dfpt} calculations, i.e. excluding the \gls{ml} component.
This \gls{rmse} is computed by determining the difference between the Hubbard parameters at the current $N_\mathrm{iter}$th and the previous $(N_\mathrm{iter}-1)$th iteration.
It is evident that the reference \gls{rmse} decreases non-monotonically, which was observed in previous studies,~\cite{Timrov:2021} and eventually diminishes at large $N_\mathrm{iter}$.
It is noteworthy that the optimal $N_\mathrm{iter}$ for achieving self-consistency in the \gls{scf} protocol varies for different materials listed in \cref{tab:materials}.
Therefore, for larger $N_\mathrm{iter}$ values in \cref{fig:iteration_comparison}, fewer data points are available compared to smaller $N_\mathrm{iter}$ values.

The \gls{ml}-based \gls{rmse} in \cref{fig:iteration_comparison} at the $N_\mathrm{iter}$th iteration is computed by training the model on data from all previous $N_\mathrm{iter}-1$ iterations and predicting the Hubbard parameters for the $N_\mathrm{iter}$th iteration.
Section~S7 in the \gls{si} contains parity plots for various values of $N_\mathrm{iter}$, ranging from 2 to 7, alongside the reference results broken down by element as well as the results for the training set.
It is evident from \cref{fig:iteration_comparison} that the \gls{ml}-based \gls{rmse} steadily decreases, and for all $N_\mathrm{iter}$ values except $N_\mathrm{iter}=6$, it is smaller than the \gls{rmse} of the reference dataset.
This result shows that our \gls{ml} model can improve upon a \gls{dfpt}-based \gls{scf} result that was terminated early, particularly during the first few iterations, thereby facilitating faster convergence of Hubbard parameters.
For instance, the reference value for $N_\mathrm{iter}=2$ indicates that on average, a single-shot \gls{dfpt} calculation yields Hubbard parameters with an \gls{rmse} of $\sim 1$~eV, while the \gls{ml} model trained on single-shot data can reduce this error to under $0.8$~eV.
Incorporating additional $N_\mathrm{iter}$ iterations in the \gls{ml} training data leads to steady improvements, although the relative performance compared to the reference gradually diminishes until reaching the floor of the model's accuracy.
This floor is partly determined by the convergence of the underlying \gls{dfpt}-based Hubbard parameters in the \gls{scf} protocol, conducted with a tolerance of approximately $\Delta = 0.01 - 0.1$~eV (see \cref{fig:self-consistent}).
Therefore, the most significant computational savings in predicting the \gls{scf} Hubbard parameters can be achieved using our \gls{ml} model, which acts as a surrogate, after only a few iterations in the \gls{dfpt}-based \gls{scf} protocol.
This substantially reduces the overall computational cost and renders it a feasible approach, particularly for high-throughput screening scenarios where \textit{a~priori} training data is unavailable for all materials.

\paragraph{Transferability of the ML model}
To test the transferability of the model, we isolate training data from the olivines (Li$_x$$M$PO$_4$ with $M$ = Fe, Mn, or Fe$_\text{0.5}$Mn$_\text{0.5}$) for which we have calculations at Li concentrations $x = 0, 0.25, 0.5, 0.75, 1$.
    Within this class of materials, the \gls{os} of \gls{tm} elements Mn and Fe changes from $+2$ to $+3$ upon delithiation.~\cite{Cococcioni:2019}
    As mentioned earlier, this change of \gls{os} is directly reflected in changes to the occupation matrices.~\cite{Timrov:2022b}
    Consequently, the Hubbard parameters for \gls{tm} elements at different \gls{os} also exhibit variation.
    For instance, the \gls{scf} Hubbard $U$ parameter for Mn changes from 4.56 to 6.26~eV when transitioning from $+2$ to $+3$, while for Fe this change is from 5.29 to 5.43~eV.~\cite{Timrov:2022b}
    In addition, in Sec.~S8 of the \gls{si} we present the distribution of Hubbard $U$ parameters for Fe and Mn ions in the olivines extracted from the \gls{dfpt}-based \gls{scf} protocol for various Li concentrations.
    Our \gls{ml} model effectively captures these changes in the Hubbard parameters and accurately predicts their values based on the occupation matrices for each TM ion at different Li concentrations.

\begin{figure*}[ht]
    \centering
    \includegraphics[width=\textwidth]{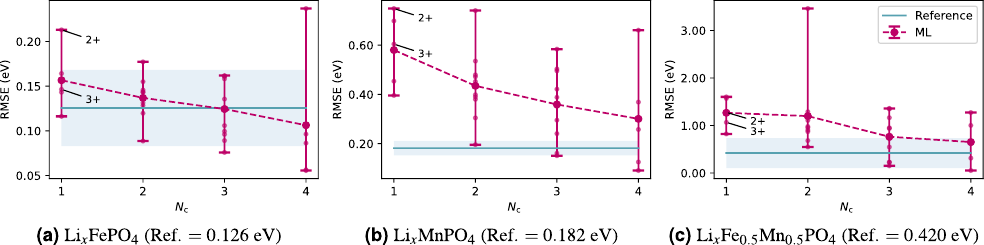}
    \caption{\textbf{Evaluation of model transferability to unseen electronic structures.}
        RMSE as a function of $N_\mathrm{c}$ (the number of Li concentrations) for three olivines.
        The model is trained using $N_\mathrm{c}$ concentrations and validated on the remaining $5-N_\mathrm{c}$.
        Error bars indicate the range of RMSE values obtained by considering various permutations of concentrations of the training and validation datasets, large dots represent the average, small dots show individual results and experiments that only contain training data from a single oxidation state are labelled ($2+$ and $3+$).
        The reference is the RMSE computed by training the ML model on 80\% of the Hubbard parameters from all five concentrations, with validation performed on the remaining 20\% after de-duplication.
        The blue line represents the mean results over three runs with different random initializations, the confidence interval shows the standard deviation.}
    \label{fig:predict_olivines}
\end{figure*}

    The outcomes of our numerical experiments are shown in \cref{fig:predict_olivines}.
    As a reference, we compute the \gls{rmse} individually for each material by training the \gls{ml} model on 80\% of the data from all five concentrations and validating it on the remaining 20\% of the data after de-duplication.
    This yields \gls{rmse} values of 126, 182, and 420~meV for the Fe, Mn, and mixed Fe-Mn olivines, respectively (see the horizontal lines in \cref{fig:predict_olivines}).
    Subsequently, we investigate the \gls{rmse} values for scenarios in which the \gls{ml} model is trained on fewer concentrations, denoted as $N_\mathrm{c}$, to assess the sensitivity of the \gls{ml} model to the amount of the training data and its transferability for predicting Hubbard parameters at other concentrations.
    To accomplish this, we train the \gls{ml} model on $N_\mathrm{c}$ concentrations and validate it on the remaining $5-N_\mathrm{c}$ concentrations, and then we average out over all possible permutations of the concentrations between the training and validation datasets.
    The average \gls{rmse} values are represented by larger dots, smaller dots show each individual result, while the error bars in \cref{fig:predict_olivines} indicate the maximum and minimum \gls{rmse} values resulting from the various permutations.
    When training on data from a single concentration (N$_c$ = 1) we have labelled the results from $x = 0$ and $1$ which correspond to 3+ and 2+ formal \glspl{os} respectively, in these cases the model is being tested on \glspl{os} it has not seen during training.
    Four out of six of these give a lower \gls{rmse} than experiments where the training data included a mixture of 2/3+ states, suggesting that the model is capable of making meaningful predictions outside of the \gls{os} seen during training.
    More generally, as $N_\mathrm{c}$ increases, the average \gls{rmse} for each material decreases, reaching values of approximately 106, 301, and 652~meV at $N_\mathrm{c} = 4$ for the Fe, Mn, and mixed Fe-Mn olivines, respectively.

    To test whether our model can transfer between different crystal structures we perform an additional experiment where we isolate data from the tunnel ($\alpha$) and rutile ($\beta$) phases of MnO$_2$, training on one and testing predictions on the other.
    \Cref{tab:structure_transferability} shows that the model can transfer well between the two phases, achieving \glspl{rmse} comparable to those found when training on 80\% of the data from both phases and predicting on the rest.

    \begin{table}
        \centering
        \begin{tabular}{lc}
            \hline\hline
            Trained on                          & RMSE (meV) \\
            \hline
            80\% of $\alpha$ \& $\beta$-MnO$_2$ & 144        \\
            $\alpha$-MnO$_2$                    & 132        \\
            $\beta$-MnO$_2$                     & 157        \\
            \hline\hline
        \end{tabular}
        \caption{Hubbard $U$ results when training on $\alpha$-MnO$_2$ data and predicting on $\beta$-MnO$_2$ and vice versa as well as a reference trained on 80\% of all of the deduplicated data and predicting on the remainder.
            In this case, the model shows good transferability to the unseen crystal structures.
        }
        \label{tab:structure_transferability}
    \end{table}

    Overall, these experiments demonstrate that, depending on the desired accuracy, it is not necessary to include training data from the exact target system that the model is being evaluated on, and even using examples of one or two similar structures may be sufficient to extrapolate to other compositions or crystal structures.
    Given the computational cost of \gls{scf} calculations, this can lead to a significant speedup, particularly in cases requiring larger supercells and various configurations.
    These findings demonstrate the good transferability of our \gls{ml} model, indicating its potential utility and reliability for predicting Hubbard parameters in materials not included in the model's training dataset.



    \section*{DISCUSSION}
    We have introduced a novel equivariant \gls{ml} model designed for predicting the self-consistent on-site $U$ and inter-site $V$ Hubbard parameters, thereby circumventing the computationally intensive \gls{dfpt}-based protocols.
    The model incorporates three input descriptors: Hubbard parameters, inter-atomic distances, and, notably, atomic occupation matrices.
    The latter play a pivotal role in encoding essential information about the electronic structure and local chemical environment within materials.
    Such an \gls{ml} model holds significant promise, particularly for high-throughput investigations and large-scale systems, scenarios where \gls{dfpt}-based approaches become too computationally expensive.
    Furthermore, the model demonstrates good transferability, rendering it reliable for predicting Hubbard parameters in materials not included in its training dataset, or as a very accurate first guess for further \gls{dfpt} refinements.

    The usage of our \gls{ml} model is straightforward and entails two \gls{dft}-based calculations for a given material.
    Initially, a ground-state calculation employing DFT+$U$+$V$ with initial guesses for $U$ and $V$ (which can be set to zero) is conducted to determine the atomic occupation matrices required as input for the model.
    Subsequently, a final structural optimization using the model-predicted \gls{scf} Hubbard parameters yields a self-consistent structural-electronic ground state.
    The computational cost of these two calculations and a model evaluation is negligible compared to the \gls{dfpt}-based \gls{scf} protocol they replace (see Sec.~S10 in the SI).
    Furthermore, we have demonstrated that the full \gls{scf} protocol can be reduced to just a few iterations, which prove adequate for achieving a \gls{rmse} within a few percent of the real value.  

    Unlike other \gls{ml} models designed for predicting Hubbard parameters,~\cite{Yu:2020, Cai:2024, Tavadze:2021, Yu:2023} our model does not rely on experimental data or information from other state-of-the-art computational methods such as $GW$ and hybrid functionals, which possess inherent limitations and specific ranges of applicability.
    Instead, our model exclusively relies on linear-response theory through \gls{dfpt}, which provides material-specific Hubbard parameters directly reflecting the local chemistry and \gls{os} of \gls{tm} elements. Additionally, the current ML model can be trained on different spin configurations of the same material, with these variations reflected in the occupation matrices, though the resulting changes in Hubbard parameters are minor~\cite{Mahajan:2021, Mahajan:2022}.
    Notably, our model not only predicts on-site Hubbard $U$ parameters but also inter-site Hubbard $V$ parameters, crucial for materials characterized by significant covalent interactions, an aspect that was disregarded in previous studies.~\cite{Discussion_v}
    Furthermore, the architecture of our \gls{ml} model is highly versatile, permitting easy integration of additional inputs and outputs, facilitating exploration of diverse learning tasks beyond learning Hubbard parameters.
    We expect that similar hybrid \gls{ml}-accelerated electronic structure methodologies, maintaining accuracy and transferability, will become prevalent, potentially yielding a comparable impact on the field as observed with \gls{ml} interaction potentials.

    Lastly, it is essential to highlight the limitations of the trained models we have presented.
    These were trained on data generated using a specific computational setup (see below), which must remain exactly the same when the model is applied to other systems.
    In other words, the $U$ and $V$ values predicted by the model are not transferable across electronic-structure codes and even across different pseudopotentials within the same code.~\cite{Kulik:2008}
    Furthermore, the training data only encompass Fe, Mn, and Ni with Hubbard $U$ corrections, while $V$ parameters are available for pairs involving these elements and O.
    Consequently, these models can effectively predict Hubbard parameters for other materials with Hubbard corrections on these atoms or atom pairs, provided that the model input is generated using identical pseudopotentials, Hubbard projectors, and functional (further elaborated in the methods section).
    Nevertheless, with a more diverse dataset, the models could be easily extended to accommodate a broader range of compositions.
    Indeed, ongoing efforts aim to establish a comprehensive database of Hubbard parameters for various \gls{tm}-containing materials, akin to Ref.~\citenum{Moore:2024}, thereby significantly broadening the scope of our model to encompass numerous \gls{tm} elements across diverse \gls{os} and chemical environments.
    While we have shown that the models can predict the results of fully self-consistent parameters, models which predict the output of single \gls{dfpt} calculations may be more applicable in certain situations (see Sec.~S11 in the SI).
    We have shown that models can be trained for this task but have not explicitly tested their performance under recursive evaluation, i.e. performing the self-consistent convergence procedure replacing \gls{dfpt} completely with an ML model.
    The performance of the models in this use case and the potential for active learning during this process could be investigated in future work.
    Moreover, our model can integrate into automated AiiDA workflows,~\cite{Huber2020, Uhrin2021} enabling non-experts to harness it effortlessly and access \gls{scf} Hubbard parameters with minimal intervention.
    Consequently, we believe that our \gls{ml} model represents a significant advancement in expediting materials discovery, design, and understanding based on the DFT+$U$+$V$ approach, thereby unlocking new avenues for technological progress and breakthroughs.


    \section*{METHODS}

    \subsection*{\textit{Ab initio} calculations}

    All calculations are performed using the plane-wave pseudopotential method as implemented in the \QE\ distribution.~\cite{Giannozzi:2009, Giannozzi:2017, Giannozzi:2020}
    We use the exchange-correlation functional constructed using \gls{sgga} with the PBEsol prescription,~\cite{Perdew:2008} and the pseudopotentials are taken from the SSSP library~v1.1 (efficiency).~\cite{Prandini:2018, MaterialsCloud}
    For metallic ground states, we use Gaussian smearing.
    To construct the Hubbard projectors, we use atomic orbitals which are orthonormalized using L\"owdin's method.~\cite{Lowdin:1950, Mayer:2002}
    Structural optimizations are performed using DFT+$U$+$V$~\cite{Timrov:2020b} with the Broyden-Fletcher-Goldfarb-Shanno (BFGS) algorithm~\cite{Fletcher:1987} and convergence thresholds for the total energy of $10^{-6}$~Ry, for forces of $10^{-5}$~Ry/Bohr, and for pressure of $0.5$~Kbar.
    The \gls{dfpt} calculations of Hubbard parameters are performed using the \textsc{HP} code,~\cite{Timrov:2022} with an accuracy of $0.01-0.1$~eV for the computed values of $U$ and $V$.
    The information about the kinetic-energy cutoff, $\mathbf{k}$ and $\mathbf{q}$ points sampling of the Brillouin zone for each system are detailed in Sec. S9 in the SI.
More technical details can be found directly in the source files publicly available through the Materials Cloud Archive (see below).


\section*{DATA AVAILABILITY}

The data used to generate the results presented in this paper are accessible in the Materials Cloud Archive.~\cite{MaterialsCloudArchive2024}


\section*{CODE AVAILABILITY}

The latest version of the code for the equivariant neural networks is available at \url{https://github.com/camml-lab/hubbardml/}.
Results presented in this paper were generated using version 0.2.0 \cite{martin_uhrin_2024_13942029}, which is also available on the Materials Cloud Archive.~\cite{MaterialsCloudArchive2024}


\section*{ACKNOWLEDGEMENTS}
We thank Francesco Aquilante, Mario Geiger, and Jigyasa Nigam for fruitful discussions.
We acknowledge support by the NCCR MARVEL, a National Centre of Competence in Research, funded by the Swiss National Science Foundation (Grant number 205602). This work was supported by a grant from the Swiss National Supercomputing Centre (CSCS) under project ID~s1073 (Piz Daint) and ID~465000416 (LUMI-G).
This work has been partially supported by MIAI@Grenoble Alpes, (ANR-19-P3IA-0003).


\section*{Author contributions statement}

M.U. developed and trained \gls{ml} model architecture; A.Z. wrote data processing code;  L.B. and I.T. performed self-consistent calculations of Hubbard parameters; M.U., A.Z., L.B., N.M. and I.T. analysed the results and wrote the manuscript.

\section*{Competing interests}

The authors declare no competing interests.








\bibliography{biblio}

\clearpage

\end{document}


\maketitle

\section{Hubbard parameters from DFPT}

In Hubbard-corrected DFT, the values of Hubbard parameters are not known {\it apriori}, making first-principles calculations of Hubbard parameters essential and highly desirable. We compute $U$ and $V$ through a generalized piecewise linearity condition imposed via linear-response theory,~\cite{Cococcioni:2005} based on DFPT.~\cite{Timrov:2018, Timrov:2021} Within this framework, the Hubbard parameters are defined as:
\begin{equation}
    U^I = \left(\chi_0^{-1} - \chi^{-1}\right)_{II} \,,
    \label{eq:Ucalc}
\end{equation}
and
\begin{equation}
    V^{IJ} = \left(\chi_0^{-1} - \chi^{-1}\right)_{IJ} \,,
    \label{eq:Vcalc}
\end{equation}
where $\chi_0$ and $\chi$ are the bare and self-consistent susceptibilities, measuring the response of atomic occupations to shifts in the potential acting on individual Hubbard manifolds. $\chi$ is defined as
\begin{equation}
    \chi_{IJ} = \sum_{m\sigma} \frac{dn^{I \sigma}_{mm}}{d\alpha^J} \,,
\end{equation}
where $n^{I\sigma}_{m m'} \equiv n^{II\sigma}_{m m'}$ is a short-hand notation for the onsite occupation matrix, $\alpha^J$ is the strength of the perturbation of electronic occupations of the $J$th site, and it is computed at self-consistency of the DFPT calculation. $\chi_0$ has a similar definition, but it is computed before the self-consistent re-adjustment of the Hartree and exchange-correlation potentials.~\cite{Timrov:2018} The response of the occupation matrix is computed in a unit cell as:
\begin{equation}
    \frac{dn^{I \sigma}_{mm'}}{d\alpha^J} = \frac{1}{N_{\mathbf{q}}}\sum_{\mathbf{q}}^{N_{\mathbf{q}}} e^{i\mathbf{q}\cdot(\mathbf{R}_{l} - \mathbf{R}_{l'})}\Delta_{\mathbf{q}}^{s'} n^{s \sigma}_{mm'} \,,
    \label{eq:dnq}
\end{equation}
where $\mathbf{q}$ is the wavevector of the monochromatic perturbation, $N_\mathbf{q}$ is the total number of perturbations, $\Delta_{\mathbf{q}}^{s'} n^{s \sigma}_{mm'}$ is the lattice-periodic response of atomic occupations to a $\mathbf{q}$-specific monochromatic perturbation. $I\equiv(l,s)$ and $J\equiv(l',s')$, where $s$ and $s'$ are the atomic indices in unit cells while $l$ and $l'$ are the unit cell indices, $\mathbf{R}_l$ and $\mathbf{R}_{l'}$ are the Bravais lattice vectors. The quantities $\Delta_{\mathbf{q}}^{s'} n^{s \sigma}_{mm'}$ are computed from the response Kohn-Sham wavefunctions, obtained by solving $\mathbf{q}$-specific Sternheimer equations. More details about the DFPT approach can be found in Refs.~\citenum{Timrov:2018, Timrov:2021}. The $\mathbf{q}$-point mesh must be chosen dense enough to make the atomic perturbations decoupled from their periodic replicas. It is important to recall that the main advantage of DFPT over the traditional linear-response approach\cite{Cococcioni:2005} is that it does not require the usage of computationally expensive supercells. Finally, it is crucial to remind that the values of the computed Hubbard parameters strongly depend on the type of Hubbard projector functions $\phi^I_{m}(\mathbf{r})$ used in the definition of the occupation matrix [see Eq.(3) in the main text] and the Hubbard potential. In this work, we use the atomic orbitals orthogonalized using the L\"owdin method.~\cite{Lowdin:1950, Mayer:2002}

\section{Self-consistent protocol}

Here we explain in more detail the self-consistent protocol for computing Hubbard parameters which is illustrated in Fig.~1(a) in the main text.
The process is initiated by the selection of the crystal structure, which can be taken from crystal structure databases, and an initial guess for the input Hubbard parameters, $U_\mathrm{in}$ and $V_\mathrm{in}$, which can be set to zero.
Following this, a structural optimization is performed using DFT+$U_\mathrm{in}$+$V_\mathrm{in}$, encompassing Hubbard forces and stresses.~\cite{Timrov:2020b}
Subsequently, a DFPT calculation is performed on the relaxed ground state to obtain the output values of Hubbard parameters, $U_\mathrm{out}$ and $V_\mathrm{out}$. If the input and output Hubbard parameters, as well as the geometry, differ beyond user-specified thresholds, the procedure iterates by updating the input Hubbard parameters and geometry. This iterative cycle continues until convergence is achieved for both the Hubbard parameters and the crystal structure.
At the end of this iterative procedure, the final \gls{scf} Hubbard parameters, $U_\mathrm{SC}$ and $V_\mathrm{SC}$, are obtained. These parameters are then used for production calculations using DFT+$U_\mathrm{SC}$+$V_\mathrm{SC}$. It is worth noting that the \gls{scf} cycle can take other forms. For instance, one may keep the geometry fixed and converge the Hubbard parameters through multiple DFPT calculations.
Only then is the structure updated, followed by the convergence of Hubbard parameters, and this iteration continues until the \gls{scf} solution is reached.
The self-consistent protocol is crucial as it guides the system toward the ground state where the electronic structure and crystal structure are mutually consistent. Additionally, it is important to mention that this protocol can be adapted for use in the DFT+$U$ framework by setting inter-site $V$ to zero.

Figure~1(b) in the main text provides an example of applying the self-consistent protocol to LiMnPO$_4$ in the framework of DFT+$U$.~\cite{Timrov:2022b} In the first iteration, the Hubbard $U$ parameter for Mn-$3d$ states is approximately 5.08~eV, computed on top of the \gls{sgga} (PBEsol)~\cite{Perdew:2008} ground state (i.e. $U_\mathrm{in} = 0$). However, by the end of the self-consistent protocol, the \gls{scf} value for $U$ is 4.1~eV — a significant change compared to a ``single-shot'' calculation (i.e., only the first iteration). The observed change is approximately 1~eV, showing a substantial and impactful adjustment for production calculations. It is important to stress that the output $U$ value after the first iteration strongly depends on the guess of the input $U$ value. Hence, even larger variations may be observed between $U$ values after the first and final iterations. The inter-site $V$ values follow a similar convergence trend.~\cite{Timrov:2021} Numerous studies have demonstrated that DFT+$U$+$V$ calculations with \gls{scf} values for $U$ and $V$ exhibit remarkable agreement with experiments across various materials and properties.~\cite{Ricca:2020, Floris:2020, Mahajan:2021, Zhou:2021, Mahajan:2022, Timrov:2022b, Timrov:2023, Binci:2023}
The self-consistent workflows in our database take on average 4 steps to converge the Hubbard parameters with an accuracy of $\Delta = 0.01-0.1$~eV, suitable for the majority of applications.

\section{Spherical harmonic conventions}
To be able to use the occupation matrices as calculated by \gls{qe} it is necessary to perform a change of basis from their convention to that used by \texttt{e3nn}, the neural network library that we use for generating equivariant learnable functions.
The following differences need to be accounted for:
\begin{enumerate}
    \item \gls{qe} uses the Condon-Shortley phase convention i.e. the spherical harmonics are defined as
          \begin{equation}
              Y_{\ell}^m={
                      \begin{cases}
                          \left(-1\right)^{m}{\sqrt {2}}{\sqrt {{\dfrac {2\ell +1}{4\pi }}{\dfrac {(\ell -|m|)!}{(\ell +|m|)!}}}}\;P_{\ell }^{|m|}(\cos \theta )\ \sin(|m|\varphi ) & {\text{if }}m   < 0  \\
                          {\sqrt {\dfrac {2\ell +1}{4\pi }}}\ P_{\ell }^{m}(\cos\theta )                                                                                            & {\text{if }}m=0      \\
                          \left(-1\right)^{m}{\sqrt {2}}{\sqrt {{\dfrac {2\ell +1}{4\pi }}{\dfrac {(\ell -m)!}{(\ell +m)!}}}}\;P_{\ell }^{m}(\cos \theta )\ \cos(m\varphi )         & {\text{if }}m   > 0.
                      \end{cases}
                  }
          \end{equation}

          while in \texttt{e3nn} the factor of $(-1)^m$ is not absorbed into the definition of $Y_\ell^m$,
    \item \gls{qe} uses the convention that for vectors ($\ell = 1$) $Y_1^{-1} = p_y$, $Y_1^0 = p_z$, $Y_1^1 = p_x$ while \texttt{e3nn} uses $Y_1^{-1} = p_x$, $Y_1^0 = p_y$, $Y_1^1 = p_z$, and,
    \item \gls{qe} prints values in the order $m = 0, 1, -1, \ldots, l, -l$ while \texttt{e3nn} expects $m = -\ell, -\ell + 1, \ldots, 0, \ldots, \ell - 1, \ell$.\footnote{In addition, \gls{qe} uses column-vector format as opposed to row-vector as used by Python, however this does not affect the occupation matrices as they are symmetric.}
\end{enumerate}
For the case of $\ell = 1$ the change of basis from \gls{qe} to \texttt{e3nn} is simply the permutation-reflection matrix
\begin{equation}
    Q^{\prime1}_{m^\prime m} = \mathcal{P}_{m^\prime m}^1 Q_{m^\prime m}^1 =
    \begin{bmatrix}
        -1           & 0 & \phantom{-}0 \\
        \phantom{-}0 & 1 & 0            \\
        \phantom{-}0 & 0 & -1
    \end{bmatrix}
    \begin{bmatrix}
        0 & 0 & 1 \\
        1 & 0 & 0 \\
        0 & 1 & 0
    \end{bmatrix}
\end{equation}
However for $\ell > 1$ the situation is more complicated as the spherical harmonics at a given $\ell$ are no longer related by a simple permutation (e.g. $d_{xy}$, $d_{yz}$, $d_z^2$, $d_{xy}$ and $d_{x^2 - y^2}$).
For these cases, we can use the, so called, Wigner-D matrices:
\begin{equation}
    D_{m\prime m}^l (\alpha, \beta, \gamma) = \braket{ lm^\prime | \hat{R}(\alpha, \beta, \gamma) | lm }
\end{equation}
where $\hat{R}(\alpha, \beta, \gamma) = e^{-i\alpha J_z}e^{-i\beta J_y}e^{-i\gamma J_z}$ is the rotation operator parameterised by the three Euler angles.
By plugging in $Q_1$ we can get the appropriate rotation matrix at any angular frequency, $Q^\ell = D_{m^\prime m}^\ell (Q_1)$, which can then be pre-multiplied by $\mathcal{P}^\ell = (-1)^{m} \delta_{m^\prime m}$ to get any $Q^{\prime\ell}$.
Finally, to make \gls{qe}'s occupation matrices compatible with \texttt{e3nn} we simply perform the transformation:
\begin{equation}
    n_{m^\prime m}^{\prime\ell} = Q_{m^\prime m}^{\prime\ell} n_{m^\prime m}^{\ell} (Q_{m^\prime m}^{\prime\ell})^T.
\end{equation}



\section{De-duplication procedure}

Our ML model treats each Hubbard active site (or pair of sites) as independent inputs.
However, a unit cell may contain more than one symmetry-equivalent site leading to near-identical inputs and outputs.
This would bias our statistics when evaluating the performance of the model, particularly if a random split of the data ends up putting two or more near-identical inputs into the training and validation splits, respectively, as the model will be trained on data that it is being validated on.
To avoid this, before each training run we de-duplicate the data by finding clusters of near-identical inputs and choosing only one example which then (depending on the nature of the numerical experiment) ends up in either the training or validation set.

\begin{figure}[t]
    \begin{subfigure}[b]{0.98\textwidth}
        \includegraphics[width=\textwidth]{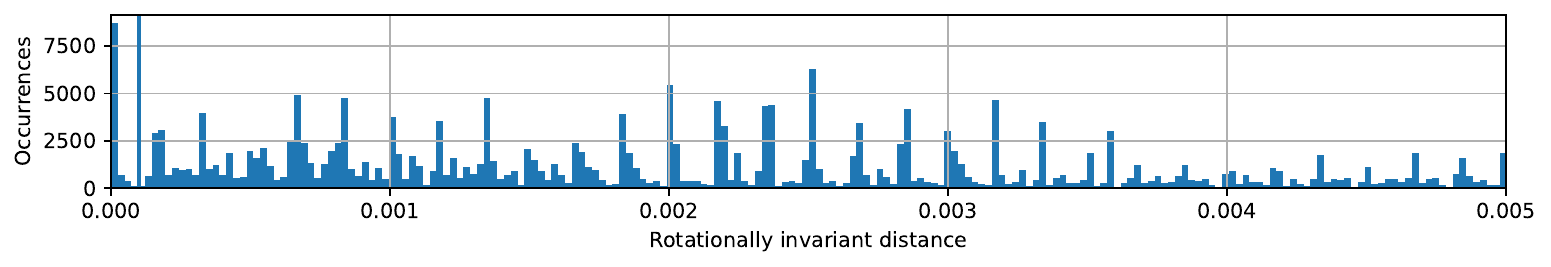}
        \caption{Distance analysis for $\textbf{x}_\ell^1$}
    \end{subfigure}\\
    \begin{subfigure}[b]{0.98\textwidth}
        \includegraphics[width=\textwidth]{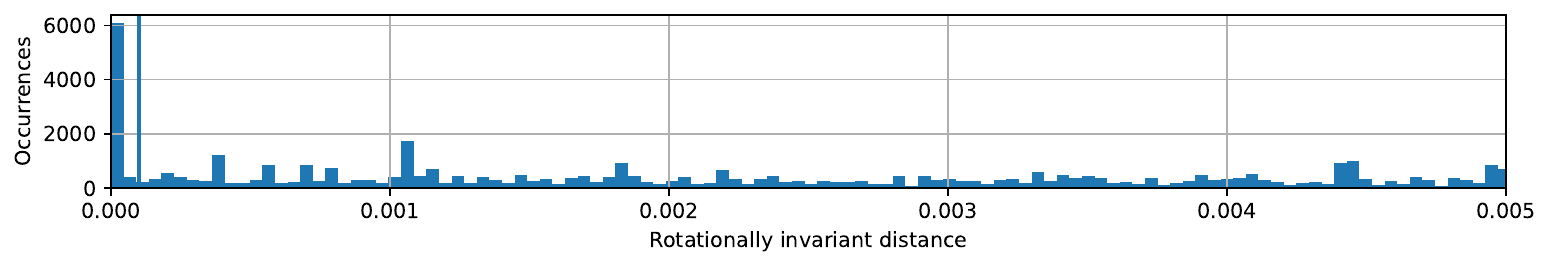}
        \caption{Distance analysis for $\textbf{x}_\ell^2$}
    \end{subfigure}
    \caption{Analysis of the symmetry-invariant distance distribution for the two permutationally invariant tensors, $\mathbf{x}^1_{\ell}$ and $\mathbf{x}^2_{\ell}$.  The vertical line represents the chosen threshold for defining two occupation matrices as identical.}
    \label{fig:occs_distance_threshold}
\end{figure}

To compare permutationally invariant matrices $\mathbf{x}^i_{\ell}$, we calculate the power spectrum distances between pairs of sites:
\begin{equation}
    d^{i, IJ}_\ell = \sum_k \left(\mathbf{x}^i_{\ell}\right)_k^I \cdot \left(\mathbf{x}^i_{\ell}\right)_k^J \,,
\end{equation}
where $i$ can be 1 or 2 (see Eqs.~(4) and (5) in the main text), $\ell$ is the angular momentum of the manifold, $k$ labels each irrep tensor in the spherical harmonic basis, while $I$ and $J$ label the atomic sites.
We calculate $d^{i, IJ}_\ell$ for all pairs of inputs (grouped by species) and form clusters from all data points that are within the thresholds defined in~\cref{tab:thresholds}.
The similarity threshold for power spectrum distances was chosen by considering the histogram of distances plotted in \cref{fig:occs_distance_threshold} and choosing a value just above the large peak near zero.
This peak should capture the identical electronic occupations with the minimum to its right separating it from the meaningfully distinguishable configurations.


\begin{table}[h!]
    \renewcommand{\arraystretch}{1.3}
    \begin{center}
        \begin{tabular}{l l}
            \hline\hline
            Attribute        & Similarity threshold   \\
            \hline
            $U_\mathrm{in}$  & $10^{-3}$ eV           \\
            $V_\mathrm{in}$  & $10^{-3}$ eV           \\
            $d^{1, IJ}_\ell$ & $10^{-4}$              \\
            $d^{2, IJ}_\ell$ & $10^{-4}$              \\
            $r_{IJ}$         & $4 \times 10^{-3}$ \AA \\  
            \hline\hline
        \end{tabular}
        \caption{Similarity thresholds for each of the input attributes provided to the ML model. $U_\mathrm{in}$ and $V_\mathrm{in}$ are the input on-site and inter-site Hubbard parameters, respectively, $\mathbf{x}^1_{\ell}$ and $\mathbf{x}^2_{\ell}$ are the tensors defined in Eqs.~(5) and (6) in the main text, and $r_{IJ}$ is the interatomic distance between species $I$ and $J$.}
        \label{tab:thresholds}
    \end{center}
\end{table}

\section{Relative error distributions}

\begin{figure}[h!]
    \centering
    \begin{subfigure}[t]{0.42\textwidth}
        \includegraphics[width=0.99\textwidth]{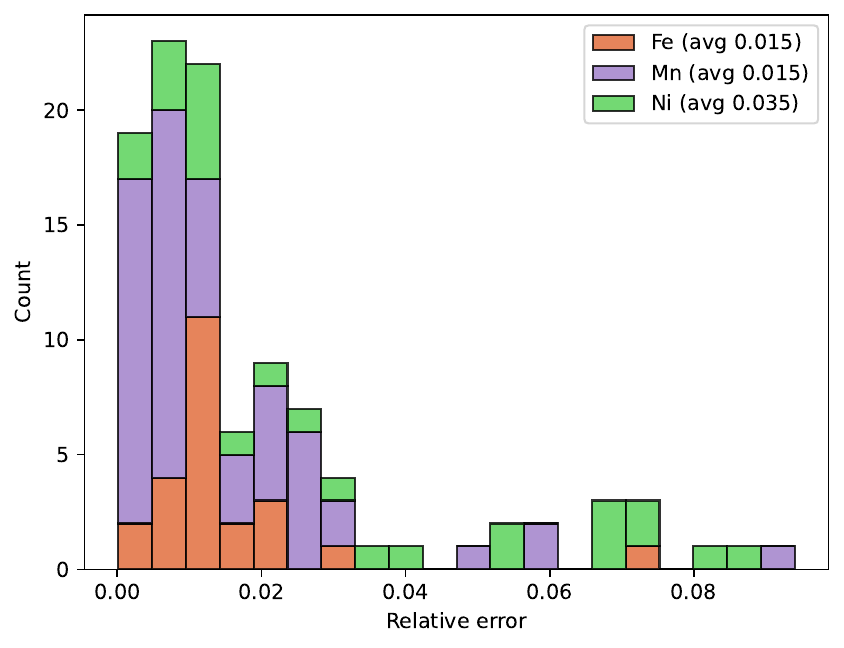}
        \caption{Hubbard $U$}
        \label{fig:hubbard_u_relative_error_validate}
    \end{subfigure}
    \begin{subfigure}[t]{0.42\textwidth}
        \includegraphics[width=0.99\textwidth]{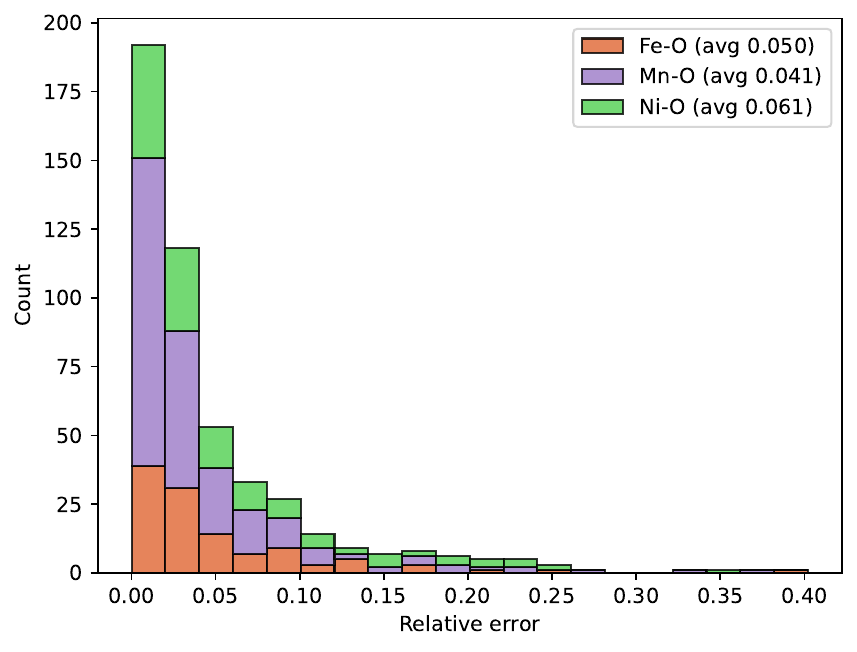}
        \caption{Hubbard $V$}
        \label{fig:hubbard_v_relative_error_validate}
    \end{subfigure}
    \caption{Validation results for predicting final SC Hubbard parameters plotted as a histogram of errors relative to the values from DFPT.}
    \label{fig:relative_errors_validate}
\end{figure}

\Cref{fig:relative_errors_validate} shows the results for predicting \gls{scf} Hubbard parameters relative to the values from DFPT. Taking Hubbard $U$ as an example, this is calculated as
$\sum_i^N \left| U_\mathrm{out}^i - \tilde{U}_\mathrm{out}^i \right| / U_\mathrm{out}$,
where the sum runs over all $N$ validation results and the predicted value is indicated with a tilde.
Given that the range of Hubbard parameters can vary depending on the element, this gives us a consistent way to assess the global performance of the model.
For both Hubbard $U$ and $V$ the trend is similar, with Fe and Mn showing similar mean relative errors, while that of Ni is consistently higher.
This may be due to the fact that Ni has a relatively wide range of Hubbard parameters (compared to Fe), but we have less training data than for Mn, which also has a much larger spread of self-consistent parameter values.

\newpage
\clearpage

\section{ML model without Hubbard parameters as inputs}

\begin{figure}[h!]
    \centering
    \begin{subfigure}[t]{0.35\textwidth}
        \centering
        \includegraphics[width=0.99\textwidth]{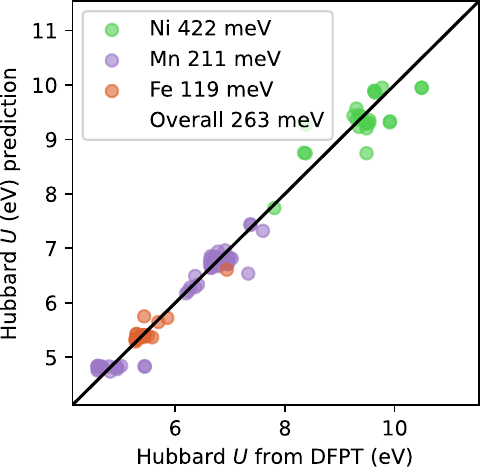}
        \caption{Hubbard $U$ (validation)}
        \label{fig:u_predict_final_no_in_param-parity_species}
    \end{subfigure}
    \hspace{1cm}
    \begin{subfigure}[t]{0.36\textwidth}
        \centering
        \includegraphics[width=0.99\textwidth]{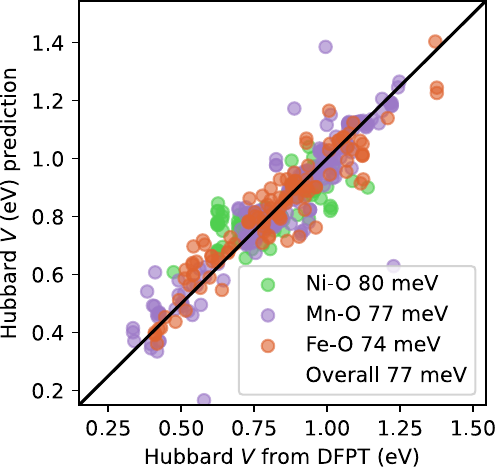}
        \caption{Hubbard $V$ (validation)}
        \label{fig:v_predct_final_no_in_param-parity_species}
    \end{subfigure}
    \caption{Parity plots showing the prediction accuracy on an unseen validation dataset, where the energies in the legend are the RMSE categorized by element(s) and the overall RMSE across all elements. All attributes listed in Table~1 in the main text (except the input Hubbard parameters) are used as inputs for the ML model. (a) Hubbard $U$ for 3d states of TM elements, (b) Hubbard $V$ between the TM-3d and O-2p states.}
    \label{fig:predict_final_no_input_param}
\end{figure}

Figure~\ref{fig:predict_final_no_input_param} shows the parity plots using our ML model when excluding the input Hubbard parameters from the list of input attributes. This numerical test demonstrates that the Hubbard parameters may not be the most critical input attributes for our ML model.

\section{RMSE and parity plots at different number of iterations $N_\mathrm{iter}$}

\begin{figure}[h!]
    \centering
    \includegraphics[width=0.5\textwidth]{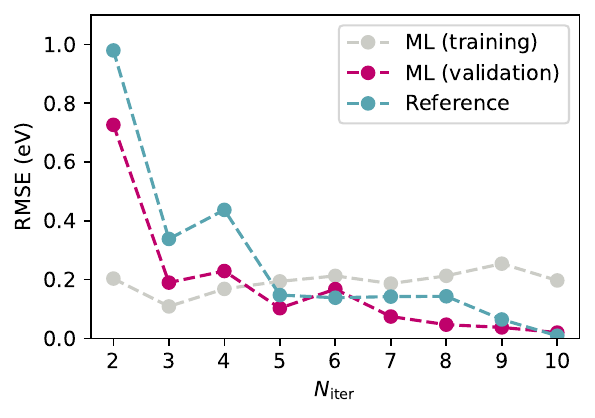}
    \caption{RMSE as a function of the number of iterations $N_\mathrm{iter}$ in the \gls{scf} protocol. The reference data represents the RMSE obtained from successive DFPT calculations (without ML involvement), whereas the ML data denotes the RMSE resulting from ML predictions for the $N_\mathrm{iter}$th iteration based on training the model on all preceding $N_\mathrm{iter} - 1$ iterations, with validation conducted using the DFPT data for the $N_\mathrm{iter}$th iteration as a reference.}
    \label{fig:iteration_comparison_w_train}
\end{figure}

\Cref{fig:iteration_comparison_w_train} show results plotted in fig. 4 of the main article, with the addition of results from the training set.
As can be seen, the RMSE on the training results never reach zero as we use early stopping to terminate training when the validation loss starts to increase.
In contrast, the validation \gls{rmse} gradually at higher $N_\mathrm{iter}$ as there are fewer and fewer materials with unconverged Hubbard parameters.

Shown below are parity plots corresponding to a model that was trained on the first $N_\mathrm{iter} - 1$ linear-response calculations, and asked to predict the Hubbard $U$ value for the current iteration $N_\mathrm{iter}$, the ``target'' is always the actual results from performing the linear-response calculation.
The reference plots use the LR result from the previous iteration as the ``prediction''.
In this way, we can assess if the model can learn to improve upon the result obtained from a limited number of DFPT calculations.

\begin{figure}[h!]
    \centering
    \begin{subfigure}[b]{\textwidth}
        \centering
        \begin{subfigure}[b]{0.48\textwidth}
            \centering
            \includegraphics[width=\textwidth]{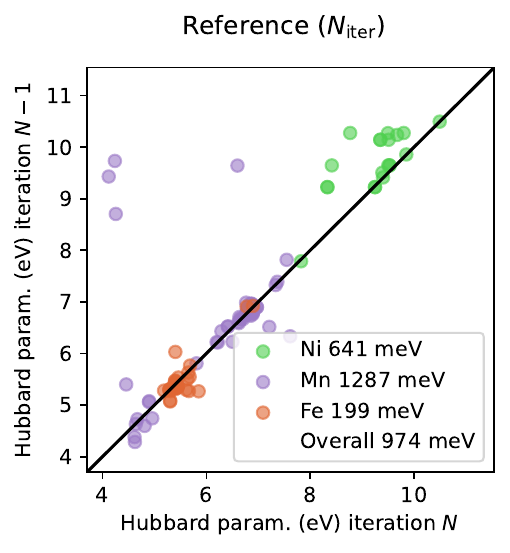}
        \end{subfigure}
        \begin{subfigure}[b]{0.48\textwidth}
            \centering
            \includegraphics[width=\textwidth]{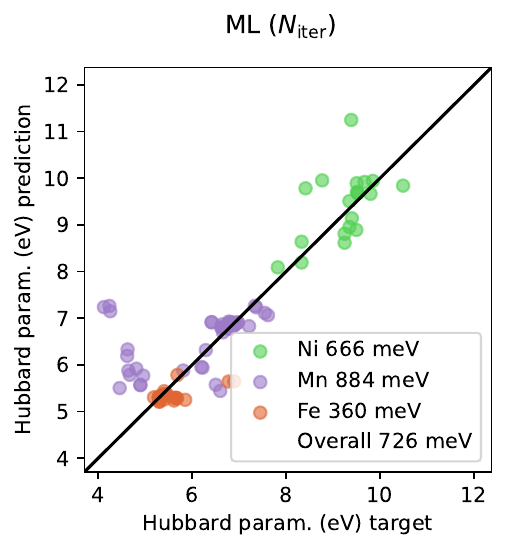}
        \end{subfigure}

    \end{subfigure}
    \caption{Second linear response step.}
\end{figure}
\begin{figure}[h!]
    \begin{subfigure}[b]{\textwidth}
        \centering
        \begin{subfigure}[b]{0.48\textwidth}
            \centering
            \includegraphics[width=\textwidth]{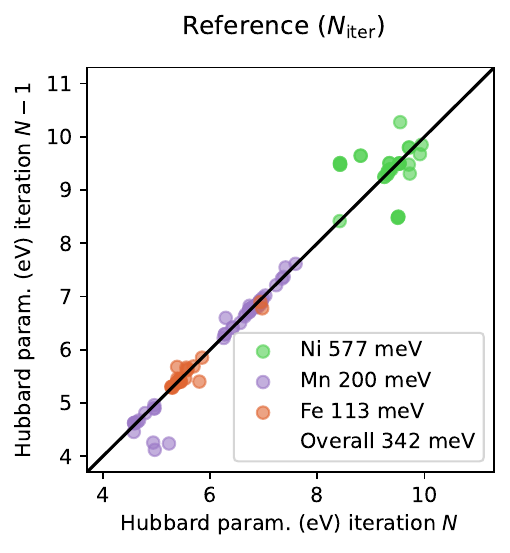}
        \end{subfigure}
        \begin{subfigure}[b]{0.48\textwidth}
            \centering
            \includegraphics[width=\textwidth]{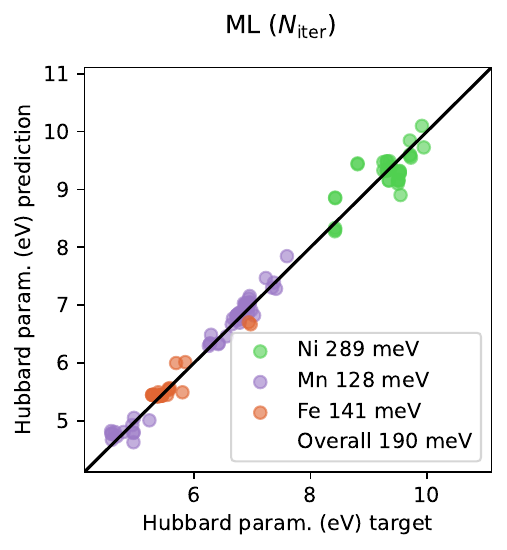}
        \end{subfigure}
    \end{subfigure}
    \caption{Third linear response step.}
\end{figure}
\begin{figure}[h!]
    \begin{subfigure}[b]{\textwidth}
        \centering
        \begin{subfigure}[b]{0.48\textwidth}
            \centering
            \includegraphics[width=\textwidth]{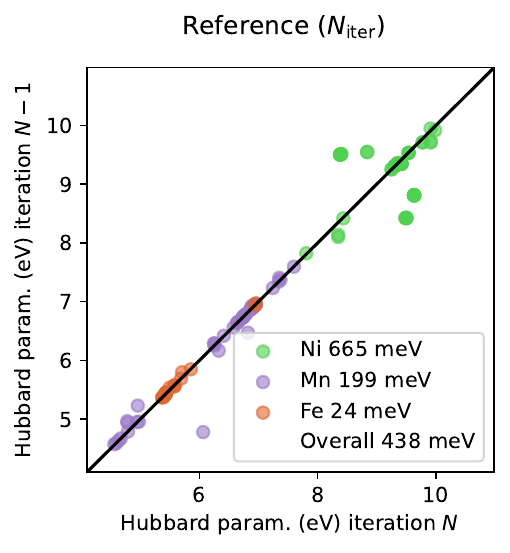}
        \end{subfigure}
        \begin{subfigure}[b]{0.48\textwidth}
            \centering
            \includegraphics[width=\textwidth]{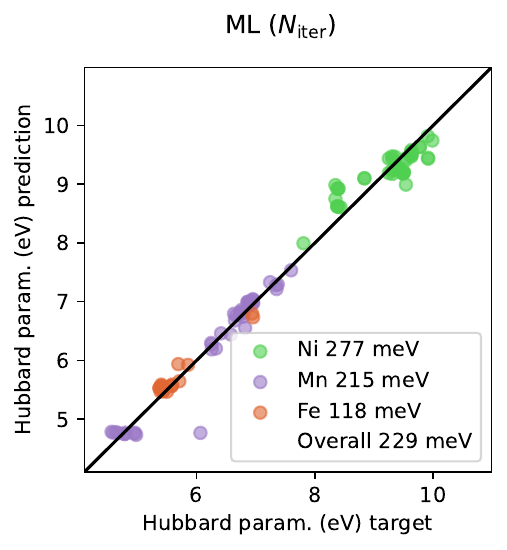}
        \end{subfigure}
    \end{subfigure}
    \caption{Fourth linear response step.}
\end{figure}
\begin{figure}[h!]
    \begin{subfigure}[b]{\textwidth}
        \centering
        \begin{subfigure}[b]{0.48\textwidth}
            \centering
            \includegraphics[width=\textwidth]{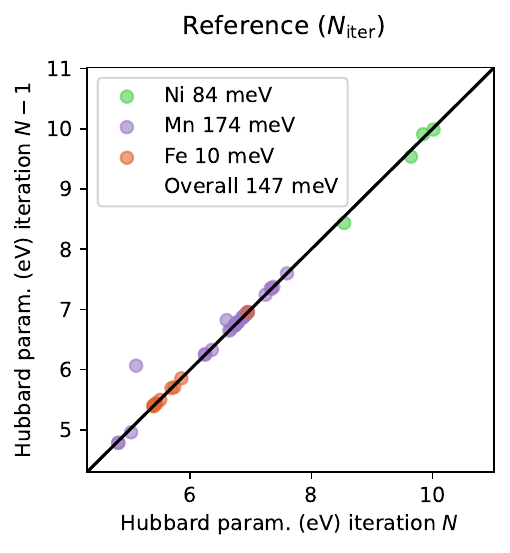}
        \end{subfigure}
        \begin{subfigure}[b]{0.48\textwidth}
            \centering
            \includegraphics[width=\textwidth]{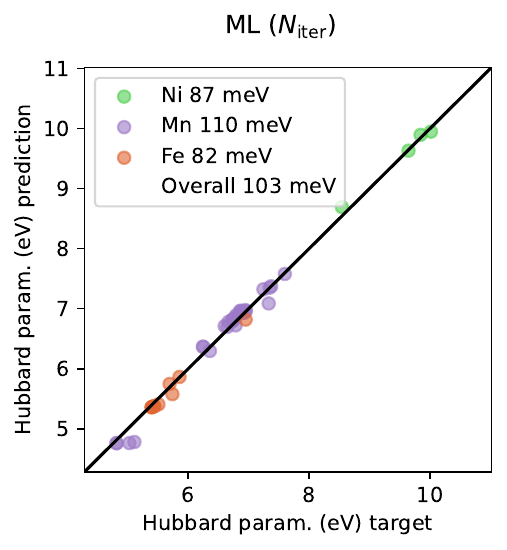}
        \end{subfigure}
    \end{subfigure}
    \caption{Fifth linear response step.}
\end{figure}
\begin{figure}[h!]
    \begin{subfigure}[b]{\textwidth}
        \centering
        \begin{subfigure}[b]{0.48\textwidth}
            \centering
            \includegraphics[width=\textwidth]{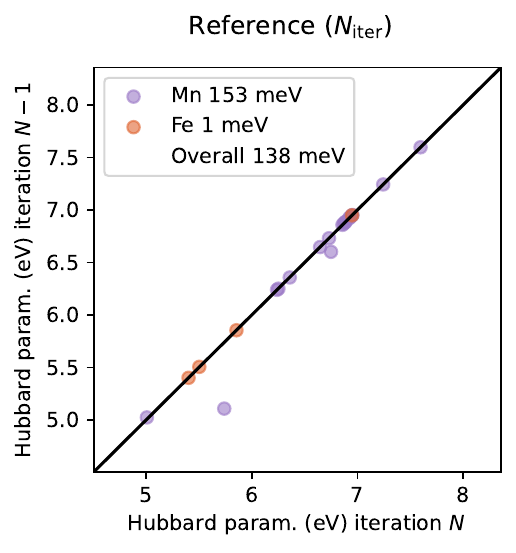}
        \end{subfigure}
        \begin{subfigure}[b]{0.48\textwidth}
            \centering
            \includegraphics[width=\textwidth]{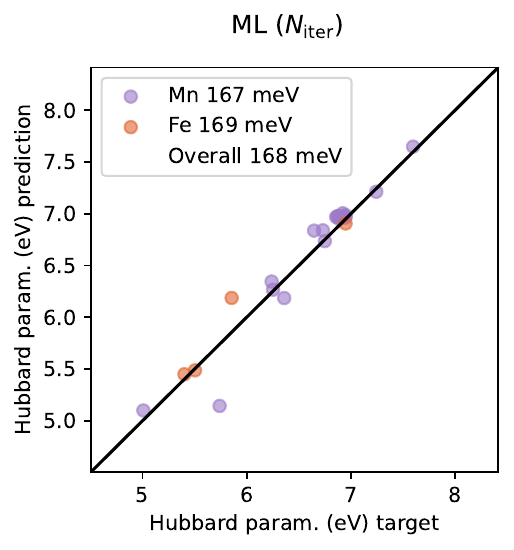}
        \end{subfigure}
    \end{subfigure}
    \caption{Sixth linear response step.}
\end{figure}
\begin{figure}[h!]
    \begin{subfigure}[b]{\textwidth}
        \centering
        \begin{subfigure}[b]{0.48\textwidth}
            \centering
            \includegraphics[width=\textwidth]{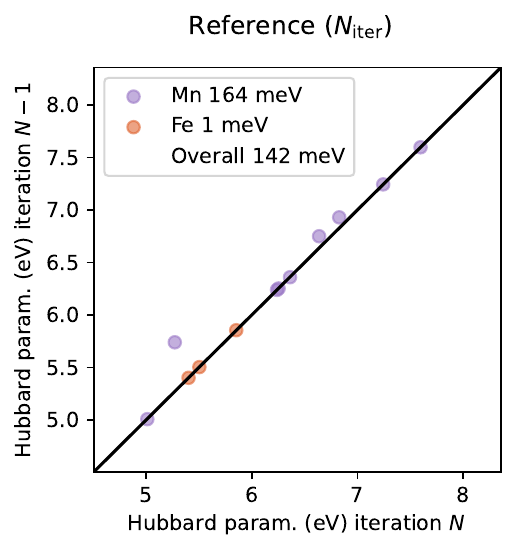}
        \end{subfigure}
        \begin{subfigure}[b]{0.48\textwidth}
            \centering
            \includegraphics[width=\textwidth]{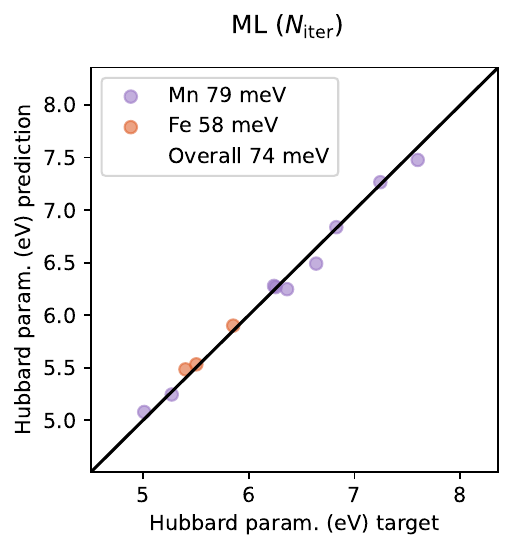}
        \end{subfigure}
    \end{subfigure}
    \caption{Seventh linear response step.}
\end{figure}

\clearpage
\section{Olivines linear-response Hubbard $U$ distribution}
\begin{figure}[h!]
    \centering
    \includegraphics[width=0.7\textwidth]{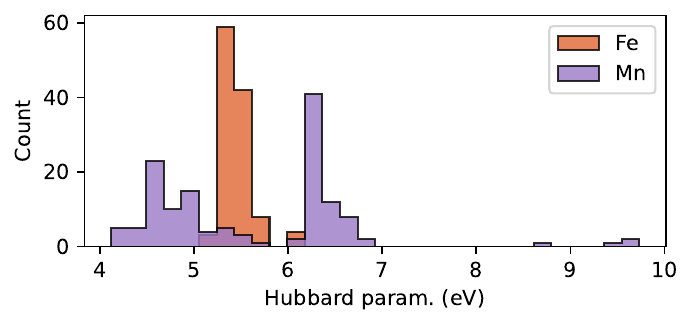}
    \caption{Distribution of Hubbard $U$ values obtained from DFPT calculations during a self-consistent procedure. Notably, the distribution of Mn parameters varies significantly more than that of Fe, with two distinct peaks corresponding to the $+2$ and $+3$ oxidation states.}
\end{figure}

\section{Additional computational details}

\begin{table}[h!]
    \centering
    \begin{tabular}{l | c  | c | c | c | c | c }
        \hline\hline
        \multirow{2}{*}{Material}        & \multirow{2}{*}{Number of atoms} & \multicolumn{2}{c|}{Structural optimization}        & \multicolumn{3}{c}{DFPT calculation of Hubbard parameters}                                                                                                       \\ \cline{3-7}
                                         &                                  & $E_\mathrm{cut}^\psi$ / $E_\mathrm{cut}^\rho$  (Ry) & $\mathbf{k}$ mesh                                          & $E_\mathrm{cut}^\psi$ / $E_\mathrm{cut}^\rho$  (Ry) & $\mathbf{k}$ mesh     & $\mathbf{q}$ mesh     \\
        \hline
        Li$_x$FePO$_4$                   & $24-28$                          & 90/1080                                             & $5 \times 10 \times 10$                                    & 65/780                                              & $3 \times 4 \times 5$ & $1 \times 2 \times 3$ \\
        Li$_x$MnPO$_4$                   & $24-28$                          & 90/1080                                             & $5 \times 10 \times 10$                                    & 65/780                                              & $3 \times 4 \times 5$ & $1 \times 2 \times 3$ \\
        Li$_x$Fe$_{0.5}$Mn$_{0.5}$PO$_4$ & $24-28$                          & 90/1080                                             & $5 \times 10 \times 10$                                    & 65/780                                              & $3 \times 4 \times 5$ & $1 \times 2 \times 3$ \\
        \hline
        Li$_x$Mn$_2$O$_4$                & $48-56$                          & 90/1080                                             & $6 \times 6 \times 6$                                      & 65/780                                              & $4 \times 4 \times 4$ & $2 \times 2 \times 2$ \\
        Li$_x$Mn$_{1.5}$Ni$_{0.5}$O$_4$  & $48-56$                          & 90/1080                                             & $6 \times 6 \times 6$                                      & 65/780                                              & $4 \times 4 \times 4$ & $2 \times 2 \times 2$ \\
        \hline
        Li$_x$NiO$_2$                    & $3 - 4$                          & 90/1080                                             & $18 \times 18 \times 10$                                   & 65/780                                              & $8 \times 8 \times 4$ & $4 \times 4 \times 2$ \\
        Li$_x$MnO$_2$                    & $3 - 4$                          & 90/1080                                             & $18 \times 18 \times 10$                                   & 65/780                                              & $8 \times 8 \times 4$ & $4 \times 4 \times 2$ \\
        \hline
        $\alpha$-MnO$_2$                 & $24-25$                          & 90/1080                                             & $4 \times 4 \times 12$                                     & 60/720                                              & $2 \times 2 \times 6$ & $1 \times 1 \times 3$ \\

        $\beta$-MnO$_2$                  & $48$                             & 90/1080                                             & $4 \times 4 \times 6$                                      & 60/720                                              & $2 \times 2 \times 4$ & $1 \times 1 \times 2$ \\
        \hline
        YNiO$_3$                         &                                  &                                                     &                                                            &                                                     &                       &                       \\
        ~~b-type                         & 40                               & 50/400                                              & $5 \times 8 \times 6$                                      & 50/400                                              & $5 \times 8 \times 6$ & $1 \times 2 \times 2$ \\
        ~~ferro                          & 40                               & 50/400                                              & $8 \times 8 \times 6$                                      & 50/400                                              & $8 \times 8 \times 6$ & $2 \times 2 \times 2$ \\
        ~~s-type                         & 40                               & 50/400                                              & $5 \times 8 \times 6$                                      & 50/400                                              & $5 \times 8 \times 6$ & $1 \times 2 \times 2$ \\
        ~~s-type-true                    & 40                               & 50/400                                              & $5 \times 8 \times 6$                                      & 50/400                                              & $5 \times 8 \times 6$ & $1 \times 2 \times 2$ \\
        ~~t-type                         & 40                               & 50/400                                              & $5 \times 8 \times 6$                                      & 50/400                                              & $5 \times 8 \times 6$ & $1 \times 2 \times 2$ \\
        ~~t-type-true                    & 40                               & 50/400                                              & $5 \times 8 \times 6$                                      & 50/400                                              & $5 \times 8 \times 6$ & $1 \times 2 \times 2$ \\
        PrNiO$_3$                        &                                  &                                                     &                                                            &                                                     &                       &                       \\
        ~~b-type                         & 80                               & 50/400                                              & $4 \times 8 \times 3$                                      & 50/400                                              & $4 \times 8 \times 3$ & $1 \times 2 \times 1$ \\
        ~~ferro                          & 20                               & 50/400                                              & $8 \times 8 \times 6$                                      & 50/400                                              & $8 \times 8 \times 6$ & $2 \times 2 \times 2$ \\
        ~~s-type                         & 80                               & 50/400                                              & $4 \times 8 \times 3$                                      & 50/400                                              & $4 \times 8 \times 3$ & $1 \times 2 \times 1$ \\
        ~~t-type                         & 80                               & 50/400                                              & $4 \times 8 \times 3$                                      & 50/400                                              & $4 \times 8 \times 3$ & $1 \times 2 \times 1$ \\
        \hline\hline
    \end{tabular}
    \caption{A list of the materials that are used to train and validate the ML model, together with the kinetic-energy cutoff for the wavefunctions ($E_\mathrm{cut}^\psi$) and charge density ($E_\mathrm{cut}^\rho$), the sizes of the $\mathbf{k}$- and $\mathbf{q}$-point meshes used for the structural optimization and DFPT calculation of Hubbard parameters.
        Multiple sizes of the $\mathbf{k}$- and $\mathbf{q}$-point meshes may have been used to accelerate calculations at the beginning of the self-consistent cycle; only the parameters used in the final iteration are reported here.}
    \label{tab:materials_comput_details}
\end{table}

\newpage

\section{Computational cost}

An important aspect of computing Hubbard parameters from first principles is computational cost.
For illustration purposes, \cref{tab:limnpo4_nodetime} reports the time cost of the variable-cell relaxation (vc-relax), DFT+$U$+$V$, and DFPT calculations performed in reaching self-consistency of the Hubbard parameters of Li$_{0.25}$MnPO$_4$.
It is worth noting that this timing is highly sensitive to the convergence parameters of the DFT+$U$+$V$ and DFPT calculations among a multitude of other factors; we list this information with the purpose of evaluating the relative cost of the different types of calculations in our highly converged study.
For this material, vc-relax and DFPT calculations of the Hubbard parameters require approximately 200$\times$ and 400$\times$ the computational resources of equivalent DFT+$U$+$V$ calculations, respectively. The vc-relax calculation is much more computationally expensive than the DFT+$U$+$V$ ground-state calculation because the former not only requires multiple ionic and cell optimization steps, but also calculations of forces and stresses containing Hubbard contributions. These Hubbard contributions are especially costly when using L\"owdin-orthonormalized atomic orbitals as Hubbard projectors~\cite{Timrov:2020b}.

\begin{table}[h!]
    \centering
    \begin{tabular}{l|rrrrrrrr|r}
    \toprule
    Iteration & 1 & 2 & 3 & 4 & 5 & 6 & 7 & 8 & Total \\
    \midrule
    vc-relax (node hours) & 236 & 359$^\dag$ & 279 & 229 & 216 & 152 & 105 & 57 & 1,633 \\
    DFT+$U$+$V$ (node hours) & 0.85 & 0.99 & 1.00 & 0.87 & 0.93 & 0.94 & 0.91 & 0.98 & 7.47 \\
    DFPT (node hours) & 402 & 430 & 389 & 424 & 413 & 419 & 416 & 414 & 3,307  \\
    \midrule
    Total (node hours) & 639 & 790 & 669 & 654 & 630 & 572 & 522 & 472 & 4,948 \\
    \bottomrule
    \end{tabular}
    \caption{Time cost in node hours for each step of the self-consistent Hubbard workflow for Li$_{0.25}$MnPO$_4$. $^\dag$The variable-cell relaxation for iteration 2 was stopped early due to non-convergence.}
    \label{tab:limnpo4_nodetime}
\end{table}

\begin{figure}[h!]
    \centering
    \includegraphics[width=0.75\textwidth]{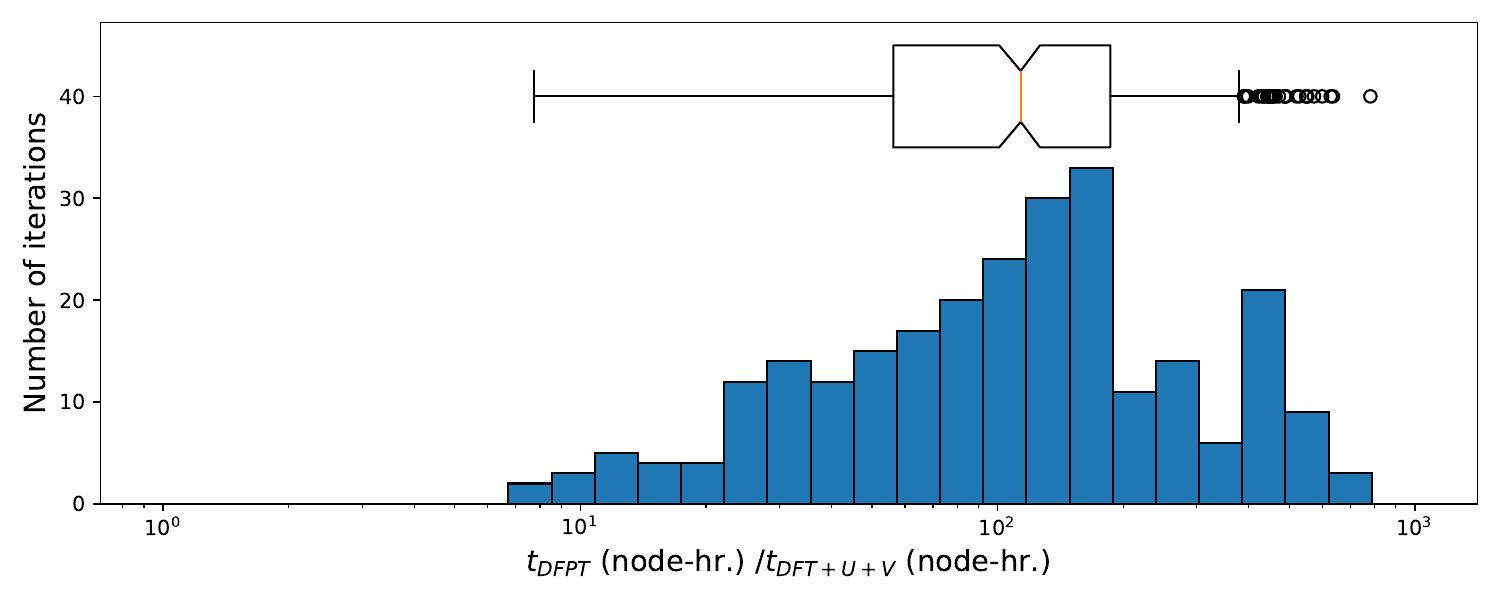}
    \caption{Ratio of node-hour time costs between DFPT and DFT+$U$+$V$ calculations per iteration over the dataset. On average, the DFPT calculations cost 2 orders of magnitude more than the DFT+$U$+$V$ calculations (median time ratio of 114).}
    \label{fig:dfpt_dft_time_ratio}
\end{figure}

\Cref{fig:dfpt_dft_time_ratio} shows a histogram of the DFPT to DFT+$U$+$V$ time ratio for all the SC Hubbard iterations in our dataset, where the median value of the cost multiplier is 114$\times$.

\begin{figure}[h!]
    \centering
    \includegraphics[width=0.75\textwidth]{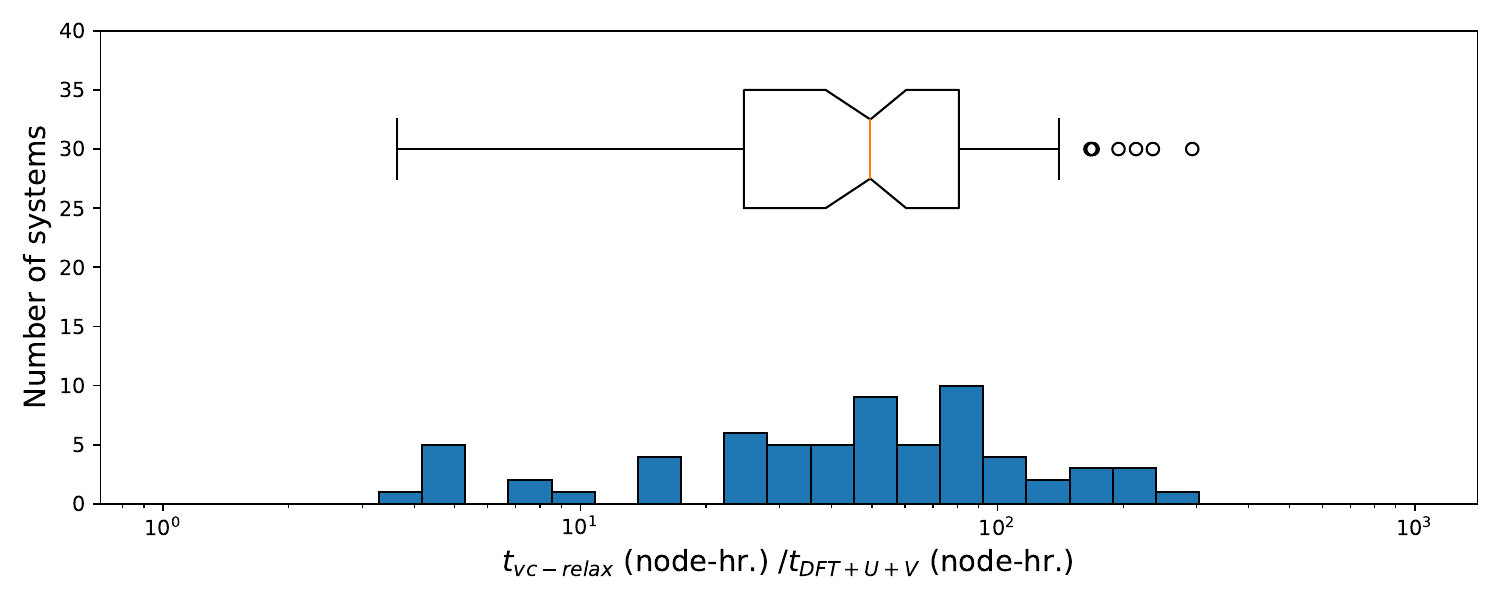}
    \caption{Ratio of node-hour time costs between the vc-relax and DFT+$U$+$V$ calculations computed per-system-mean over the dataset. On average, the vc-relax calculations cost one order of magnitude more than the DFT+$U$+$V$ calculations (median time ratio of 50).}
    \label{fig:vc_dft_time_ratio}
\end{figure}

\begin{figure}[h!]
    \centering
    \includegraphics[width=0.75\textwidth]{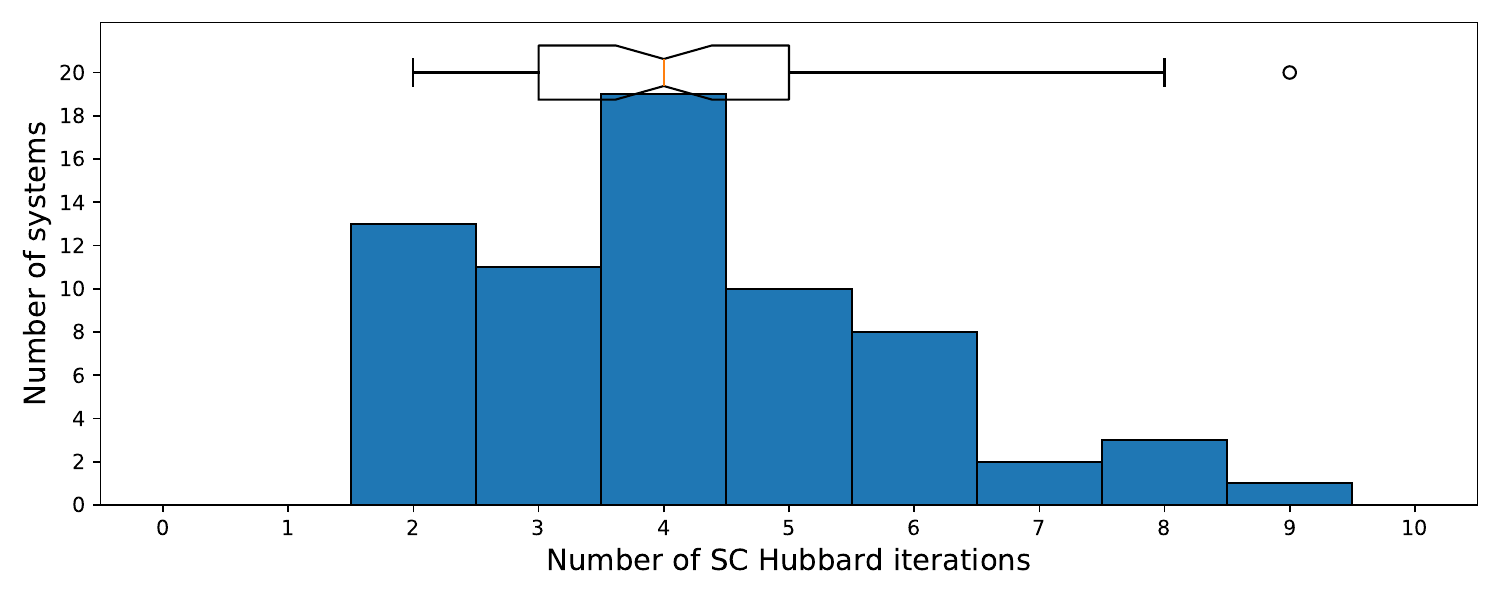}
    \caption{Number of self-consistent Hubbard iterations required to achieve convergence. Over the 67 systems represented in the database, the average number of iterations required is 4 with a standard deviation of 1.7.}
    \label{fig:sc-iters}
\end{figure}

\Cref{fig:vc_dft_time_ratio} shows the same histogram comparing vc-relax to DFT+$U$+$V$ calculations, where the median value of the cost multiplier is 50$\times$. Therefore, a median Hubbard SC iteration is expected to cost 165$\times$ that of a DFT+$U$+$V$ calculation of a given material.
In order to fully capture the average expected cost of converging the Hubbard parameters for a given material, we also report in \cref{fig:sc-iters} a histogram showing the number of iterations required for the 67 systems in our database.
The average and median number of iterations are both 4, with a standard deviation of 1.7 iterations.
Combining this with the median cost of DFPT and vc-relax calculations, it can be seen that for a median case, the expected cost of convergence is approximately 660$\pm$280 DFT+$U$+$V$ calculations.

In addition to performing DFT+$U$+$V$ and DFPT calculations to generate data, training the ML models is the other potentially significant cost associated to their use in practice.
As an example, we report here the training time for the Hubbard $U$ and $V$ models presented in the main text which are trained to predict final self-consistent Hubbard parameters.
For this study, the Hubbard $U$ model is trained in 0.85 GPU-hours and the Hubbard $V$ model in 3.73 GPU-hours.
At a total time cost of 4.58 GPU-hours, model training is $1-2$ orders of magnitude cheaper (assuming 1 consumer-grade GPU per node) than a single DFPT evaluation for a single material (measured in supercomputer CPU node-hours).

\newpage

\section{Direct substitution of DFPT calculations}

\begin{figure}[t]
    \centering
    \includegraphics[width=0.5\textwidth]{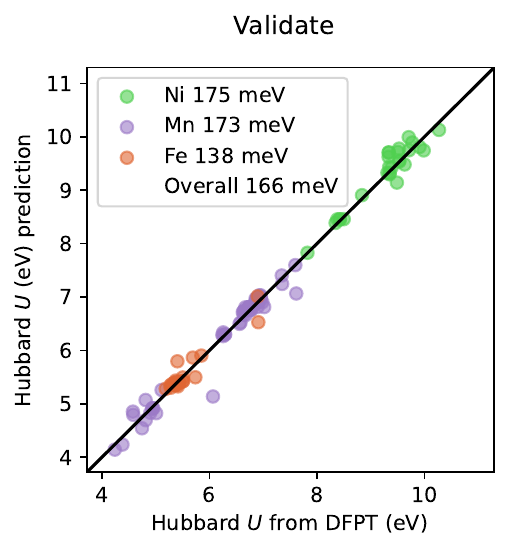}
    \caption{Parity plot showing the prediction accuracy on an unseen validation dataset when predicting the output of a DFPT calculation.}
    \label{fig:u_predict_hp_parity}
\end{figure}

In Fig.~3 of the main article we present results from a model that is trained to predict the final, self-consistent Hubbard parameters as we judge this to be the most useful way to use our model.
A somewhat easier task is to predict the outputs of a single DFPT calculation, results for which can be seen in \cref{fig:u_predict_hp_parity} when trained on Hubbard $U$ data. Comparing with fig. 3 of the main text, we see that the prediction errors on Mn and Fe are similar while for Ni the results are improved, leading to a lower overall RMSE.
By using the model in this mode it is possible to provide a direct surrogate to the DFPT calculation and carry out a self-consistent procedure, however this will incur the higher cost of having to run multiple DFT calculations, for perhaps only modest improvements in the final Hubbard parameters.
For those seeking greater accuracy (for example, as a downstream refinement step in a high-throughput study), it is likely to be more beneficial to simply use a self-consistent Hubbard parameter prediction followed by a DFT and DFPT calculations to get a refined result.

\bibliography{biblio}